\documentclass[hyper]{JHEP3} 

\usepackage{epsfig,multicol}

\newcommand\fverb{\setbox\pippobox=\hbox\bgroup\verb}
\newcommand\fverbdo{\egroup\medskip\noindent%
                        \fbox{\unhbox\pippobox}\ }
\newcommand\fverbit{\egroup\item[\fbox{\unhbox\pippobox}]}
\newbox\pippobox

\def\bea{\begin{eqnarray}}
\def\eea{\end{eqnarray}}
\def\bec{\begin{center}}
\def\ec{\end{center}}

\def\beq{\begin{equation}}
\def\eeq{\end{equation}}

\def\f{\frac}

\def\f#1#2{\frac{#1}{#2}}

\title{Supersymmetry Breaking and   Moduli Stabilization
with Anomalous U(1) Gauge Symmetry}

\author{ Kiwoon Choi $^a$ and Kwang-Sik Jeong $^{a,b}$\\
         $^a$Department of Physics, Korea Advanced Institute of Science
             and Technology\\
             Daejeon 305-701, Korea\\
         $^b$Departamento de F\'{\i}sica Te\'orica C-XI
             and Instituto de F\'{\i}sica Te\'orica  C-XVI\\
             Universidad Aut\'onoma de Madrid, Cantoblanco, E-28049 Madrid, Spain\\
         E-mail: \email{kchoi@hep.kaist.ac.kr},
                 \email{ksjeong@hep.kaist.ac.kr} }

\preprint{FTUAM 06/04\\ IFT-UAM/CSIC-06-19\\ KAIST-TH 06/07}

\abstract {We examine the effects of anomalous $U(1)_A$ gauge
symmetry on soft supersymmetry breaking terms while incorporating
the stabilization of the modulus-axion multiplet responsible for
the Green-Schwarz (GS) anomaly cancellation mechanism. In case of
the KKLT stabilization of the GS modulus, soft terms are
determined by the GS modulus mediation, the anomaly mediation and
the $U(1)_A$ mediation which are generically comparable to each
other, thereby yielding the mirage mediation pattern of
superparticle masses at low energy scale. Independently of the
mechanism of moduli stabilization and supersymmetry breaking, the
$U(1)_A$ $D$-term potential can not be an uplifting potential for
de Sitter vacuum when the  gravitino mass is smaller than the
Planck scale by many orders of magnitude. We also discuss some
features of the supersymmetry breaking by red-shifted anti-brane
which is a key element of the KKLT moduli stabilization. }


\begin{document}

\section{Introduction}

Anomalous $U(1)_A$ gauge symmetry appears often in compactified
string theory. The 4-dimensional (4D) spectrum of such compactification
contains a modulus-axion (or dilaton-axion) superfield which transforms non-linearly under
$U(1)_A$ to implement the Green-Schwarz (GS) anomaly cancellation
mechanism \cite{GS}.  In heterotic string theory, the dilaton
plays the role of the GS modulus, however in other string
theories, the GS modulus can be either
a K\"ahler modulus of Calabi-Yau (CY) orientifold \cite{BKQ,louis}
or a blowing-up modulus of orbifold singularity
\cite{ano}.
The non-linear $U(1)_A$ transformation of the
GS modulus superfield leads to a field-dependent Fayet-Iliopoulos (FI) term
\cite{anomalous} which might play an important role for
supersymmetry (SUSY) breaking. Anomalous $U(1)_A$ might also
correspond to a flavor symmetry which generates the hierarchical
Yukawa couplings through the Froggatt-Nielsen mechanism
\cite{FN0,FN}.

The $U(1)_A$ $D$-term can give a contribution to soft scalar
masses as $\Delta m_i^2=-q_ig^2_A D_A$ where $q_i$ is the $U(1)_A$
charge of the corresponding sfermion \cite{u1soft}. Such $D$-term
contribution has an important implication to the flavor problem in
supersymmetric models. If $g_A^2D_A$ is significantly bigger than
the gaugino mass-squares $M_a^2$ which are presumed to be of order
$(1 \,\,\mbox{TeV})^2$, e.g. $g_A^2D_A\sim (10\,\, \mbox{TeV})^2$,
one can avoid the SUSY flavor problem by assuming that $q_i$ are
non-vanishing only for the first and second generations of matter
fields, which would make the first and second generations of
squarks and sleptons  heavy enough to avoid dangerous
flavor-changing-neutral-current (FCNC) processes. Still one can
arrange $q_i$ to be appropriately flavor-dependent \cite{nelson}
to generate the observed pattern of hierarchical Yukawa couplings
via the Froggatt-Nielsen mechanism, e.g. $y_{ij}\sim
\epsilon^{q_i+q_j}$ for $\epsilon\sim 0.2$. In other case that
$g_A^2D_A$ is comparable to $M_a^2$, one needs $q_i$ to be
flavor-universal to avoid dangerous FCNC processes, and then
$U(1)_A$ can not be identified as a flavor symmetry for the Yukawa
coupling hierarchy. Finally, if $g_A^2D_A$ is small enough, e.g.
suppressed by a loop factor of order $10^{-2}$ compared to
$M_a^2$,  $q_i$ are again allowed to be flavor-dependent. It has
been noticed that the relative importance of the $D$-term
contribution to soft masses depends on how the GS modulus is
stabilized \cite{arkani}. In this respect, it is important to
analyze the low energy consequences of anomalous $U(1)_A$ while
incorporating  the stabilization of the GS modulus explicitly
\cite{moral,dudas,casas}.

In the previous studies of anomalous $U(1)_A$ in heterotic string
compactification, two possible scenarios for the stabilization of
the GS modulus (the heterotic string dilaton $S$ in this case)
have been considered. One is to use the multiple gaugino
condensations \cite{racetrack} which would stabilize $S$ at the
weak coupling regime for which the leading order K\"ahler
potential is a good approximation. In this race-track
stabilization,  one typically finds the auxiliary $F$ component
$F^S=0$ and also $D_A=0$, although SUSY can be broken by the
$F$-components of other moduli.
The most serious difficulty of the race-track scenario is that in
all known examples the vacuum energy density has a {\it negative}
value of  ${\cal O}(m_{3/2}^2M_{Pl}^2)$ \cite{CKN}, where
$M_{Pl}\simeq 2.4\times 10^{18}$ GeV is the 4D reduced Planck mass
and $m_{3/2}$ is the gravitino mass. Another possible scenario is
that $S$ is stabilized by (presently not calculable) large quantum
correction to the K\"ahler potential \cite{kahler}. In this case,
one can assume that the dilaton K\"ahler potential has a right
form to stabilize $S$ at a phenomenologically viable de Sitter
(dS) or Minkowski vacuum. The resulting $F^S$ and $D_A$ are
non-vanishing in general, however the relative importance of $D_A$
compared to the other SUSY breaking auxiliary components depends
sensitively on the incalculable large quantum corrections to the
K\"ahler potential \cite{arkani}.

Recently a new way of stabilizing  moduli at dS vacuum
within a controllable approximation scheme has been proposed
by Kachru-Kallosh-Linde-Trivedi (KKLT) in the context of Type IIB
flux compactification \cite{KKLT}. The main idea is to stabilize moduli (and
also the dilaton) in the first step at a supersymmetric AdS vacuum for which the leading order
K\"ahler potential is a  good approximation,
and then lift the vacuum to a dS state by adding
anti-brane.
 For instance, in Type IIB compactification, one can first
introduce a proper set of fluxes and gaugino condensations stabilizing all
moduli at SUSY AdS vacuum.
 In the next step, anti-branes can be added to get the nearly vanishing cosmological
constant under the RR charge cancellation condition. In the
presence of fluxes, the compact internal space is generically
warped \cite{GKP} and anti-branes are stabilized at the maximally
warped position \cite{KPV}. Then as long as the number of
anti-branes is small enough compared to the flux quanta,
anti-branes cause neither a dangerous instability of the
underlying compactification \cite{KPV} nor a sizable shift of the
moduli vacuum expectation values. In order to get the nearly
vanishing cosmological constant, the anti-brane energy density
should be adjusted to be close to $3m_{3/2}^2M_{Pl}^2$. This
requires that the warp factor $e^{2A}$ of the 4D metric on
anti-brane should be of ${\cal O}(m_{3/2}/M_{Pl})$. As it breaks
explicitly the $N=1$ SUSY preserved by the background geometry and
flux, one might expect that anti-brane will generate incalculable
SUSY breaking terms in the low energy effective lagrangian.
However as was noticed in \cite{choi1} and will be discussed in
more detail in this paper, the SUSY breaking soft terms in KKLT
compactification can be computed within a reliable approximation
scheme, which is essentially due to that anti-brane is red-shifted
by  a small warp factor $e^{2A}\sim m_{3/2}/M_{Pl}$.

In this paper, we wish to examine the implications of anomalous
$U(1)_A$ for SUSY breaking  while incorporating the stabilization
of the GS modulus explicitly. Since one of our major concerns is
the KKLT stabilization of the GS modulus, in section 2 we  review
the 4D effective action of KKLT compactification and discuss some
features such as the $D$-type spurion dominance and the
sequestering  of the SUSY breaking by red-shifted anti-brane which
is a key element of the KKLT compactification. In section 3, we
discuss the mass scales, $F$ and $D$ terms in generic models of
anomalous $U(1)_A$.
 In section 4, we
examine in detail a model for the KKLT stabilization of the GS
modulus and the resulting pattern of soft terms. Section 5 is the
conclusion.

The following is a brief summary of our results.
The GS modulus-axion superfield $T$  transforms under
$U(1)_A$ as \bea T\rightarrow T-i\alpha(x)\frac{\delta_{GS}}{2}\,, \eea
where $\alpha(x)$ is the $U(1)_A$ transformation function and
$\delta_{GS}$ is a constant of ${\cal O}(1/8\pi^2)$ when $T$ is
normalized as $\partial_T f_a={\cal O}(1)$ for the holomorphic
gauge kinetic functions $f_a$. There are two mass scales that
arise from the non-linear transformation of $T$:
\bea
\xi_{FI}&=& \frac{\delta_{GS}}{2}\,\partial_TK_0,
\nonumber \\
M_{GS}^2&=& \frac{\delta_{GS}^2}{4}\,\partial_T\partial_{\bar{T}}K_0,
\eea where $\xi_{FI}$ is the FI $D$-term and $M_{GS}^2$
corresponds to the GS axion contribution to the $U(1)_A$ gauge
boson mass-square
\bea
M_A^2=2g_A^2M_{GS}^2+{\cal O}(|\xi_{FI}|)
\eea
 for the K\"ahler potential $K_0$
and the $U(1)_A$ gauge coupling $g_A$. (Unless
specified, we will use the convention $M_{Pl}=1$ throughout this
paper.)  Then the $U(1)_A$ $D$-term is bounded as
\bea |D_A|\,\lesssim\, {\cal O}(m_{3/2}^2M_{Pl}^2/M_A^2)
\eea for SUSY breaking scenarios with $m_{3/2}\ll M_A$.

It has been pointed out \cite{BKQ} that the $D$-term potential
$V_D=\frac{1}{2}g_A^2D_A^2$ in models with anomalous $U(1)_A$
might play the role of an uplifting potential which compensates
the negative vacuum energy density $-3m_{3/2}^2M_{Pl}^2$ in the
supergravity potential. As the K\"ahler metric of $T$ typically
has a vacuum expectation value of order unity, we have
$M_{GS}^2\sim M_{Pl}^2/(8\pi^2)^2$. Then, since the $U(1)_A$ gauge
boson mass-square $M_A^2\gtrsim {\cal O}(M_{GS}^2)$, the above
bound on $D_A$ implies that $V_D$ is too small to be an uplifting
potential in SUSY breaking scenarios with $m_{3/2}<
M_{Pl}/(8\pi^2)^2$. In other words, models of moduli stabilization
in which $V_D$ plays the role of an uplifting potential for dS
vacuum generically predict a rather large $m_{3/2}\gtrsim {\cal
O}(M_{Pl}/(8\pi^2)^2)$ \cite{casas}. On the other hand, in view of
that the gaugino masses receive the anomaly mediated contribution
of ${\cal O}(m_{3/2}/8\pi^2)$, one needs  $m_{3/2}\lesssim {\cal
O}(8\pi^2)$ TeV
 in order to realize the supersymmetric
extension of the standard model at the TeV scale.
As a result,
models with anomalous $U(1)_A$
still need an uplifting mechanism different from the $D$-term uplifting, e.g. the anti-brane uplifting
of KKLT or a hidden matter superpotential suggested in \cite{nilles}, if $m_{3/2}$  is small enough to give the weak scale superparticle masses.

Still $D_A$ can give an important contribution to soft masses.
As we will see, the relative importance of this
$D$-term contribution   depends on the
size of the ratio \bea R\equiv \xi_{FI}/M_{GS}^2.\nonumber
\eea
If ${\rm Re}(T)$ is a string dilaton or a K\"ahler modulus which
is stabilized at a vacuum expectation value of ${\cal O}(1)$ under
the normalization $\partial_T f_a={\cal O}(1)$,
the resulting $|R|$ is of ${\cal O}(8\pi^2)$. We then find the
$D$-term contribution to soft masses is generically comparable to
the GS modulus-mediated contribution. In this case of $|R|\gg 1$,
the longitudinal component of the $U(1)_A$ gauge boson comes
mostly from the phase of $U(1)_A$ charged field $X$ with a vacuum
expectation value $\langle X\rangle\sim \sqrt{\xi_{FI}}$, rather
than from the GS axion ${\rm Im}(T)$. Then $T$ is  a
flat-direction of the $U(1)_A$ $D$-term potential, thus one needs
a non-trivial $F$-term potential to stabilize $T$. An interesting
possibility is the KKLT stabilization of $T$ involving a hidden
gaugino condensation and also anti-brane for the uplifting
mechanism. In such case, the soft terms are determined by three
contributions mediated at the scales close to $M_{Pl}$: the GS
modulus mediation \cite{KL}, the anomaly mediation \cite{AM} and
the $U(1)_A$ mediation \cite{u1soft}. Generically these three
contributions are comparable to each other, yielding the mirage
mediation pattern of superparticle masses at low energy scale
\cite{choi1,choi2,endo,choi3}. However if the K\"ahler potential
of $X$ is related to the K\"ahler potential of $T$ in a specific
manner, the $U(1)_A$ mediation is suppressed by a small factor of
${\cal O}(1/8\pi^2)$ compared to the other two mediations. Since
the anomaly mediation and the GS modulus mediation remain to be
comparable to each other, the mirage mediation pattern is
unaltered in this special case that the $U(1)_A$ mediation is
relatively suppressed.

In fact, some models of anomalous $U(1)_A$ can yield  $|R|\ll 1$.
If $T$ corresponds to a blowing-up modulus of orbifold singularity
stabilized at near the orbifold limit, one can have $|\xi_{FI}|\ll
M_{GS}^2$ \cite{ano,poppitz}, and thus  $|R|\ll 1$.
In this limit, soft terms mediated by the GS modulus at
$M_{GS}$ are negligible compared to the soft terms mediated by
a $U(1)_A$ charged field $X$  at the lower scale
$\langle X\rangle\sim \sqrt{\xi_{FI}}$. If $|R|$ is small enough, e.g. $|R|\lesssim
10^{-4}$, $U(1)_A$ $D$-term contribution is also smaller than the
low scale mediation at $\sqrt{\xi_{FI}}$.

\section{4D effective action of KKLT compactification}

In this section, we  review the 4D effective action
of KKLT compactification  and
the resulting soft SUSY breaking terms of visible fields.
We also discuss some relevant features of the SUSY breaking by
red-shifted anti-brane which is a key element of the KKLT compactification.
KKLT compactification can be split into two parts. The first part
contains the bulk of (approximate) CY space as well as the $D$
branes of visible matter and gauge fields which are assumed to be
stabilized at a region where the warping is negligible. Note that
the 4D cutoff scale  of this part should be somewhat close to
$M_{Pl}$ in order to realize the 4D gauge coupling unification at
$M_{GUT}\sim 2\times 10^{16}$ GeV. The low energy dynamics of this
part can be described by a 4D effective action which takes the
form of conventional 4D $N=1$ SUGRA: \bea S_{\rm N=1}=\int d^4x
d^2\Theta \,2{\cal E} \left[\,\frac{1}{8}({\bar{\cal D}}^2-8{\cal
R}) \left(3e^{-K/3}\right)+\frac{1}{4}f_a W^{a\alpha}W^a_\alpha
+W\,\right] +{\rm h.c.},\eea where $\Theta^\alpha$ is the Grassmann
coordinate of the curved superspace, ${\cal E}$ is the chiral
density, ${\cal R}$ is the chiral curvature superfield, and $K$,
$f_a$ and $W$ denote the K\"ahler potential, gauge kinetic
function and superpotential, respectively. In the following, we
call this part the $N=1$ sector. The scalar potential of $S_{N=1}$
in the Einstein frame is given by
\begin{eqnarray}
V_{N=1}=e^K\left\{K^{I\bar{J}}(D_I W)(D_J W)^* -3|W|^2\right\}+
\frac{1}{2{\rm Re}(f_a)} D^aD^a,
\eea
where $D_IW=\partial_IW+(\partial_IK)W$ is the K\"ahler covariant derivative
of the superpotential and $D^a=-\eta^I_a\partial_IK$ for the holomorphic Killing vector
$\eta^I_a$ of the $a$-th gauge transformation
of $\Phi^I$.
In KKLT compactification, the $N=1$ sector is assumed to have a
supersymmetric AdS vacuum\footnote{Note that
$D^a=-\eta^I_aD_IW/W$, so $D_IW=0$ leads to $D^a=0$ for $W\neq 0$.}, i.e.
\bea
\label{susyads}
\langle D_IW\rangle_{N=1}=0,\quad
\langle V_{N=1}\rangle=-3m_{3/2}^2M_{Pl}^2.
\eea

The remained part of KKLT compactification is anti-brane which
is stabilized at the end of a warped throat.
The SUSY preserved by anti-brane does not have any
overlap with the $N=1$ SUSY preserved by the background geometry
and flux.
As a consequence,
the field degrees of freedom on
anti-brane do not have $N=1$ superpartner in general.
For instance, the Goldstino fermion $\xi^\alpha$ of the broken $N=1$ SUSY
which originates from anti-brane does not have  bosonic
$N=1$ superpartner. This means  that the $N=1$ local SUSY is
non-linearly realized on the world-volume of anti-brane.
Still the anti-brane action can be written in a locally supersymmetric
superspace form
using the Goldstino superfield \cite{SW}:
\bea
\label{supergold}
\Lambda^\alpha=\xi^\alpha+\Theta^\alpha+...,
\eea
where the ellipsis denotes the $\xi^\alpha$-dependent higher order terms.
In the unitary gauge of $\xi^\alpha=0$, the anti-brane action
appears to break the $N=1$ SUSY  explicitly. Generic explicit SUSY
breaking relevant for the soft terms of visible fields is
described by three spurion operators: $D$-type spurion operator
$\tilde{\cal P}\Theta^2\bar{\Theta}^2$, $F$-type non-chiral
spurion operator $\tilde{\Gamma}\bar{\Theta}^2$, and $F$-type chiral
spurion operator  $\tilde{\cal F}\Theta^2$. Then the local
lagrangian density on the world volume of anti-brane can be
written as \bea {\cal L}_{\rm anti}&=&\delta^6(y-\bar{y})\int
d^2\Theta 2\,{\cal E}\left[\,\frac{1}{8}({\bar{\cal D}}^2-8{\cal
R}) \Big(\,e^{4A}\tilde{\cal
P}\,\Theta^2\bar{\Theta}^2+e^{3A}\tilde{\Gamma}\,\bar{\Theta}^2\,\Big)
\right.\nonumber \\
&&\left. \qquad\qquad\qquad\qquad\quad -\,e^{4A}\tilde{\cal
F}\,\Theta^2+...\,\right] +{\rm h.c.}, \eea where $\bar{y}$ is the
coordinate of the anti-brane in six-dimensional internal space,
$e^{2A}$ is the warp factor on the anti-brane world volume: \bea
ds^2(\bar{y})=e^{2A}g_{\mu\nu}dx^\mu dx^\nu, \eea and the ellipsis
stands for the Goldstino-dependent terms which are not so relevant
for us. Generically $\tilde{\cal P}$, $\tilde{\Gamma}$ and
$\tilde{\cal F}$  have a value of order unity in the unit with
$M_{Pl}=1$ (or in the unit with the string scale $M_{\rm st}=1$).
The warp factor dependence of each spurion operator can be easily
determined by noting that $\tilde{\cal P}\Theta^2\bar{\Theta}^2$
and $\tilde{\cal F}\Theta^2$ give rise to an anti-brane energy
density which is red-shifted by $e^{4A}$,
while $\tilde{\Gamma}\bar{\Theta}^2$ gives rise to a gravitino mass on the anti-brane world volume
which is red-shifted by $e^{3A}$.
(See the discussion of Appendix A for this red-shift of gravitino mass.)
Including the Goldstino fermion explicitly,
the spurion operators in ${\cal L}_{\rm anti}$ can be written in a locally supersymmetric form, e.g.
\bea
\tilde{\cal P}\Lambda^2\bar{\Lambda}^2&=& \tilde{\cal P}\Theta^2\bar{\Theta}^2+...,
\nonumber \\
\tilde{\Gamma}\bar{\Lambda}^2&=& \tilde{\Gamma}\bar{\Theta}^2+...,
\nonumber \\
\tilde{\cal F}\tilde{W}^\alpha\tilde{W}_\alpha&=&\tilde{\cal
F}\Theta^2+..., \eea where $\tilde{W}_\alpha=
\frac{1}{8}(\bar{\cal D}^2-8{\cal R}){\cal
D}_\alpha(\Lambda^2\bar{\Lambda}^2)$ and the ellipses denote the
Goldstino-dependent terms.

The SUSY breaking spurions on the world volume of anti-brane can be transmitted
to the visible $D$-branes by a bulk field propagating through the warped throat.
The warp factor dependence of spurions allows us to
estimate the size of SUSY breaking induced by each spurion without
knowing the detailed mechanism of transmission.
In addition to giving a vacuum energy density of ${\cal O}(e^{4A}M_{Pl}^4)$, the $D$-type spurion
$\tilde{\cal P}\Theta^2\bar{\Theta}^2$
can generate SUSY breaking scalar mass-squares of  ${\cal O}(e^{4A}M_{Pl}^2)$
through the effective operator $e^{4A}\tilde{\cal P}\Theta^2\bar{\Theta}^2Q^{i*}Q^i$
which might be induced by the exchange of bulk fields, where
$Q^i$ denote the visible matter superfields.
The non-chiral $F$-type spurion $\tilde{\Gamma}\bar{\Theta}^2$
might generate trilinear scalar couplings of ${\cal
O}(e^{3A}M_{Pl})$ through the effective operator $e^{3A}\tilde{\Gamma}\bar{\Theta}^2Q^{i*}Q^i$,
while the chiral $F$-type spurion $\tilde{\cal
F}\Theta^2$ might generate gaugino masses of ${\cal O}(e^{4A}M_{Pl})$ through the effective
chiral operator $e^{4A}\tilde{\cal
F}\Theta^2W^{a\alpha}W^a_\alpha$.  When combined with its
complex conjugate or with the $F$-component of $N=1$ sector
moduli, $\tilde{\Gamma}\bar{\Theta}^2$ can generate a vacuum
energy density of ${\cal O}(e^{6A}M_{Pl}^4)$ or ${\cal
O}(e^{3A}m_{3/2}M_{Pl}^3)$, and scalar mass-squares of ${\cal
O}(e^{6A}M_{Pl}^2)$ or ${\cal O}(e^{3A}m_{3/2}M_{Pl})$. Similarly,
the chiral $F$-type spurion $\tilde{\cal F}\Theta^2$ can generate
a vacuum energy density and scalar mass-squares, but they are
suppressed by one more power of $e^A$ compared to the contribution
from $\tilde{\Gamma}\bar{\Theta}^2$. In case with $e^{A}\sim 1$,
all spurions give equally important contributions of the Planck
scale size, leading to uncontrollable SUSY breaking. On the other
hand, in case that $e^{A}\sim \sqrt{m_{3/2}/M_{Pl}}$, which is in fact
required in order for that the anti-brane energy density cancels
the negative vacuum energy density (\ref{susyads}) of the $N=1$
sector, SUSY breaking terms which originate from the $F$-type spurions  are
negligible compared to the terms which originate from the $D$-type spurion
since they are suppressed by additional power of $e^A\sim
\sqrt{m_{3/2}/M_{Pl}}$.
For instance, in the presence of the $D$-type spurion providing a vacuum energy
density of ${\cal O}(m_{3/2}^2M_{Pl}^2)$,
there are always the anomaly-mediated soft masses of ${\cal O}(m_{3/2}/8\pi^2)$
which are much bigger than the soft masses induced by the $F$-type spurions
when $e^A\ll 1/8\pi^2$.
Note that  $e^A\sim\sqrt{m_{3/2}/M_{Pl}}\lesssim 10^{-6}$ for
$m_{3/2}\lesssim {\cal O}(8\pi^2)$ TeV which is necessary to
get the weak scale SUSY. Obviously, this feature of $D$-type
spurion dominance greatly simplifies the SUSY breaking by
red-shifted anti-brane.

In addition to the Goldstino fermion,
there can be other anti-brane fields, e.g. the anti-brane position moduli
$\tilde{\phi}$.\footnote{There can be also anti-brane gauge field $\tilde{A}_\mu$.
However $\tilde{A}_\mu$ is not relevant for the transmission of SUSY breaking to the visible sector,
thus will be ignored.}
The anti-brane moduli also do not have $N=1$
superpartner, however one can construct the corresponding
Goldstino-dependent superfields as
\bea \tilde{\Phi}=\tilde{\phi}+i(
\Theta\sigma^\mu\bar{\xi}-\xi\sigma^\mu\bar{\Theta})\partial_\mu\tilde{\phi}+...\eea
The anti-brane lagrangian density including $\tilde{\Phi}$ and
also the bulk moduli $\Phi$ which can have a local interaction on
the world volume of anti-brane can be written  as \bea
 \label{anti-action}
 {\cal L}_{\rm anti}&=&\delta^6(y-\bar{y})\int d^2\Theta
\,2{e^{3A}\cal E} \left[\,\frac{1}{8}e^{-A}\Big({\bar{\cal
D}}^2-8{\cal R}\Big)\Omega_{\rm anti}(Z_A,Z_A^*)\, \right]+{\rm
h.c.}, \eea where $\Omega_{\rm anti}$ is a function of
$Z_A=\Big\{\,e^{A/2}\Lambda^\alpha, e^{-A/2}{\cal D}_\alpha,
e^{-A}{\cal R}, \tilde{\Phi}, \Phi \,\Big\}$. Here the warp factor
dependence of ${\cal L}_{\rm anti}$ is determined by the Weyl
weights of the involved superfields. Taking into account that the
$F$-type spurions can be ignored in case of $e^{A}\sim
\sqrt{m_{3/2}/M_{Pl}}$, $\Omega_{\rm anti}$ can be approximated as \bea
\Omega_{\rm anti}&\simeq
&e^{2A}\Lambda^2\bar{\Lambda}^2\left[\tilde{\cal P}(\Phi,\Phi^*)
+\frac{1}{16}e^{-2A}Z_{\tilde{\Phi}}(\Phi,\Phi^*)\tilde{\Phi}^*
\bar{\cal D}^2{\cal
D}^2\tilde{\Phi}+M_{\tilde{\Phi}}^2(\Phi,\Phi^*)\tilde{\Phi}^*\tilde{\Phi}
\right], \eea where $Z_{\tilde{\Phi}}={\cal O}(1)$,
$M_{\tilde{\Phi}}={\cal O}(M_{Pl})$, and
$\langle\tilde{\Phi}\rangle$ is chosen to be zero. This shows that
the anti-brane moduli masses are generically of ${\cal
O}(\sqrt{m_{3/2}M_{Pl}})$. Since it is confined on the world
volume of anti-brane, $\tilde{\Phi}$ can not be a messenger of
SUSY breaking, so can be integrated out without affecting the
local SUSY breaking in the visible sector. Then, after integrating
out the KK modes of bulk fields as well as the anti-brane moduli $\tilde{\Phi}$,
the 4D effective action induced by $\Omega_{\rm anti}$ takes the form:
\bea  S^{(4D)}_{\rm anti} = \frac{1}{8}\int
d^4xd^2\Theta\, 2{\cal E}\,(\bar{\cal D}^2-8{\cal R})
\left(\tilde{\cal P}
(\Phi,\Phi^*)+\tilde{\cal Y}_i(\Phi,\Phi^*)Q^{i*}Q^i \right)
e^{4A}\Lambda^2\bar{\Lambda}^2+{\rm h.c.}
\eea
Note that the contact interaction between
$e^{4A}\Lambda^2\bar{\Lambda}^2$ and $Q^{i*}Q^i$ was not allowed in
$\Omega_{\rm anti}$ because $\Lambda^\alpha$ and the visible matter superfields
$Q^i$ live on different branes
which are geometrically separated from each other.
Thus, if $\tilde{\cal Y}_i\neq 0$,
it should be a consequence of
the exchange of bulk fields which couple to both $e^{4A}\Lambda^2\bar{\Lambda}^2$
(on anti-brane) and $Q^{i*}Q^i$ (on the $D$-branes of visible fields).

Possible phenomenological consequences of $S^{(4D)}_{\rm anti}$
are rather obvious. The Goldstino operator $e^{4A}\tilde{\cal
P}\Lambda^2\bar{\Lambda}^2$ gives rise to an uplifting potential
of ${\cal O}(m_{3/2}^2M_{Pl}^2)$ which would make the total vacuum
energy density to be nearly vanishing. In the following, we will
call this Goldstino operator the uplifting operator.\footnote{In
fact, this corresponds to the superspace expression of the
Volkov-Akulov Goldstino lagrangian density.} The uplifting
potential induces also a SUSY-breaking shift of the vacuum
configuration (\ref{susyads}), which would result in nonzero
vacuum values of $F^I$ and $D^a$. The effective contact
interaction between $Q^i$ and ${\Lambda}^\alpha$ gives soft
SUSY-breaking sfermion mass-squares of ${\cal O}(\tilde{\cal
Y}_im_{3/2}^2)$. Note that the features of the 4D effective action of anti-brane which
have been discussed so far rely {\it only} on that anti-brane
is red-shifted by the warp factor $e^{A}\sim \sqrt{m_{3/2}/M_{Pl}}$, thus are valid
for generic  KKLT
compactification.

Since the scalar masses from the the effective contact
term $e^{4A}\Lambda^2\bar{\Lambda}^2Q^{i*}Q^i$ in $S^{(4D)}_{\rm anti}$
can be phenomenologically important, let us consider in what
situation this contact interaction can be generated. The warped
throat in KKLT compactification has approximately the geometry of
$T_5\times{\rm AdS}_5$ where $T_5$ is a compact 5-manifold which
is topologically $S^2\times S^3$. In the limit that the radius of
$T_5$ is small compared to the length of the warped throat,
the transmission of SUSY breaking through the throat
can be described by a supersymmetric 5D Randall-Sundrum (RS) model
\cite{RS} with visible $D$-branes at the UV fixed point
($y=0$) and
anti-brane at the IR fixed point ($y=\pi$) \cite{hebecker}.
Let us thus examine the possible generation of the effective contact
term  within the framework of
the supersymmetric 5D RS model.

It has been
noticed that the  5D bulk SUGRA multiplet does not generate a
contact interaction between UV superfield and IR superfield at
tree level \cite{luty}.\footnote{In CFT interpretation, this might
correspond to the conformal sequestering discussed in
\cite{conformal}.} Loops of 5D SUGRA fields generate such contact
interaction, however the resulting coefficient $\tilde{\cal Y}_i$
is suppressed by the warp factor $e^{2A}$
\cite{rattazzi}, so is
negligible.\footnote{Note the difference of
the SUSY-breaking IR brane operator between our case and  \cite{rattazzi}.
In our case, the SUSY breaking IR brane operator is given by
$e^{4A}\Lambda^2\bar{\Lambda}^2$ for the Goldstino superfield
$\Lambda^\alpha$ normalized as (\ref{supergold}), while
the SUSY breaking IR brane operator of \cite{rattazzi} is $e^{2A}Z^*Z$
for a $N=1$ chiral IR brane superfield $Z$ with nonzero $F^Z$.}
 In fact, in order to generate the contact interaction
$e^{4A}\Lambda^2\bar{\Lambda}^2Q^{i*}Q^i$ in 4D effective action,
one needs a bulk field $B$ other than the 5D SUGRA multiplet which has
a non-derivative coupling in $N=1$ superspace to both
$e^{4A}\Lambda^2\bar{\Lambda}^2$ at the IR fixed point and $Q^{i*}Q^i$
at the UV fixed point.
Since the SUGRA multiplet is not crucial for the following discussion,
we will use the rigid $N=1$ superspace  for simplicity, and then
the required fixed point couplings of $B$ can be written as
\bea
\int d^2\theta d^2\bar{\theta} \Big[
\delta(y)g_B BQ^{i*}Q^i+\delta(y-\pi)g_B^\prime e^{4A}B\Lambda^2\bar{\Lambda}^2
+{\rm h.c.}\Big],
\eea
where $\theta^\alpha$ is the Grassmann coordinate of the rigid $N=1$ superspace.
 If $B$ is a
chiral superfield in $N=1$ superspace,
the effective contact interaction between $Q^{i*}Q^i$ and
$e^{4A}\Lambda^2\bar{\Lambda}^2$ induced by the exchange of $B$
is suppressed by the superspace derivative $\bar{\cal D}^2$.
This can be easily noticed from the fact that the effective contact interaction arises
from the part of the solution of $B$ which is proportional to the UV brane source $Q^{i*}Q^i$
or the IR brane source $e^{4A}\Lambda^2\bar{\Lambda}^2$.
Since the brane sources are non-chiral, this part of the
solution should include  the chiral projection operator $\bar{\cal D}^2$.
As a result, the coefficient of the induced contact interaction
is given by $\tilde{\cal Y}_i \sim g_Bg^\prime_B\bar{\cal D}^2/k$
where $k$ is the AdS curvature which is essentially of
${\cal O}(M_{Pl})$.  Since $\bar{\cal D}^2/k$ leads to an additional suppression by
$m_{3/2}/M_{Pl}$,
the contact interaction induced by
chiral bulk superfield gives at most a contribution of ${\cal O}(m_{3/2}^3/M_{Pl})$
to the soft scalar mass-squares of $Q^i$ when $e^{A}\sim \sqrt{m_{3/2}/M_{Pl}}$, which is totally negligible.

On the other hand, if $B$ is a vector superfield in $N=1$ superspace,
there is no such suppression by the chiral projection operator, so
the resulting $\tilde{\cal Y}_i$ can be sizable in certain cases.
To examine the contact term induced by a bulk vector superfield in more detail, one can consider
the 5D lagrangian of $B$ which contains
\bea
\int d^2\theta d^2\bar{\theta}
&\Big[&\frac{1}{8}B{\cal D}_\alpha\bar{\cal D}^2{\cal D}^\alpha B+M_B^2e^{-2kLy}B^2
\nonumber \\
&&+\,\delta(y)g_B BQ^{i*}Q^i+\delta(y-\pi)g_B^\prime e^{-4kLy}B\Lambda^2\bar{\Lambda}^2
\,\Big],
\eea
where
$e^{-kLy}$ is the position dependent warp factor in
AdS$_5$, $L$ is the orbifold length, and $M_B$ is the 5D mass of the vector superfield $B$.
($e^{-\pi kL}=e^A$ in this convention.)
The warp factor dependence of each term in the above 5D lagrangian can be determined by looking at
the dependence on the background spacetime metric.
Note that the UV brane coupling $g_B$ (the IR brane coupling $g_B^\prime$)
corresponds to the gauge coupling between the 4D vector
field component of $B$ and the 4D current component of
the UV brane operator $Q^{i*}Q^i$  (the IR brane operator $\Lambda^2\bar{\Lambda}^2$).
The 5D locality and dimensional analysis suggest that
the contact term obtained by integrating out $B$ has a coefficient
$\tilde{\cal Y}_i\propto e^{-\pi M_B L}$ in the limit $M_B\gg k$.
Indeed, for $M_B\gtrsim k$, a more careful analysis \cite{kim} gives
\bea
\tilde{\cal Y}_i\,\sim\, {g_Bg^\prime_Be^{-\pi (\sqrt{M_B^2+k^2}-k)L}}/{M_B^2}.
\eea
This result indicates that a sizable contact term can be induced
if the model contains a vector superfield $B$ propagating through the warped throat
with bulk mass $M_B\lesssim k$ and also sizable $g_B$ and $g^\prime_B$.
In  KKLT compactification
of Type IIB string theory, one does not have such bulk vector superfield, thus
it is expected that $\tilde{\cal Y}_i$ is negligibly small,
i.e. anti-brane is sequestered well from the
$D$-branes of visible fields.   In fact, in  KKLT compactification of Type
IIB string theory, one finds that $\tilde{\cal P}$ is independent of
the CY volume modulus \cite{choi1}, thus even the CY volume modulus is
sequestered from anti-brane. This is not suprising in view of that
the wavefunction of the volume modulus has a negligible value over
the throat, thus the volume modulus can be identified as a UV
brane field in the corresponding RS picture \cite{hebecker}. In
the following, we will assume that $Q^i$ and $\Lambda^\alpha$ are
sequestered from each other, thus
\bea
\tilde{\cal Y}_i=0.
\eea
We stress
that this sequestering assumption is relevant {\it only} for the
soft scalar masses. The other SUSY breaking observables such as
the gaugino masses and trilinear scalar couplings are {\it not}
affected even when $\tilde{\cal Y}_i$ has a sizable value.

According to the above discussion, the 4D effective action of anti-brane  is highly dominated
by the uplifting operator:
 \bea  S^{(4D)}_{\rm anti} \,\simeq \,\frac{1}{8}\int
d^4xd^2\Theta\, 2{\cal E}\,(\bar{\cal D}^2-8{\cal R})
\Big(\,e^{4A}\tilde{\cal
P}(\Phi,\Phi^*)\Lambda^2\bar{\Lambda}^2\,\Big) +{\rm h.c.}, \eea
and then the total 4D effective action of KKLT compactification is
given by \bea \label{superaction} S_{\rm
KKLT}&=&S_{\rm N=1}+S^{(4D)}_{\rm anti} \nonumber \\
&=& \int d^4xd^2\Theta \,2 {\cal E} \left[\,\frac{1}{8}\big(\bar{\cal
D}^2-8{\cal R}\big) \Big(\,3e^{-K/3}+{\cal
P}\Lambda^2\bar{\Lambda}^2\,\Big) \right.
\nonumber \\
&& \left.\qquad\qquad\qquad +\,\frac{1}{4}f_a
W^{a\alpha}W^a_\alpha + W \,\right]+{\rm h.c.}, \eea where \bea
{\cal P}(\Phi,\Phi^*)=e^{4A}\tilde{\cal P}(\Phi,\Phi^*)={\cal
O}(m_{3/2}^2M_{Pl}^2). \eea Since the vacuum expectation value of
${\cal P}$ can be fixed by the condition of vanishing cosmological
constant, the above 4D effective action is almost equally
predictive as the conventional $N=1$ SUGRA without the anti-brane
term ${\cal P}\Lambda^2\bar{\Lambda}^2$. This nice feature of KKLT
compactification is essentially due to that anti-brane is highly
red-shifted.

In fact, for the discussion of moduli stabilization at a nearly
flat dS vacuum and the subsequent SUSY breaking in the visible
sector, the SUGRA multiplet can be simply replaced by their vacuum
expectation values, e.g. $g_{\mu\nu}=\eta_{\mu\nu}$ and
$\psi_\mu=0$, {\it except for} its scalar auxiliary component
${\cal M}$ whose vacuum expectation value should be  determined by
minimizing the scalar potential. The most convenient formulation
for the SUSY breaking by ${\cal M}$ is to introduce the chiral
compensator superfield $C$, then choose the superconformal gauge
 ${\cal M}=0$ to trade ${\cal M}$ for $F^C$, and finally replace the SUGRA multiplet by
their vacuum values, while making the superconformal gauge choice
in the rigid superspace:
\bea C=C_0+\theta^2F^C. \eea
In the unitary gauge,
this procedure corresponds to the following replacements for the
superspace action: \bea && \Lambda^\alpha \,\rightarrow\,
\frac{C^*}{C^{1/2}}\theta^\alpha,
\nonumber \\
&& W^{a\alpha}\,\rightarrow\, C^{-3/2}W^{a\alpha},\nonumber \\
&& d^2\Theta 2{\cal E}\,\rightarrow\, d^2\theta C^3, \nonumber \\
&& -\frac{1}{4}d^2\Theta 2{\cal E}(\bar{\cal D}^2-8{\cal R}) \,\rightarrow\,
d^2\theta d^2{\bar{\theta}}CC^*,
 \eea
under which the locally supersymmetric action (\ref{superaction})
 is changed to
 \bea
 \label{rigid-action} S_{\rm KKLT}&=& \int d^4x \left[
\,\int d^2\theta d^2\bar{\theta} \,CC^*
\left(\,-3e^{-K/3}-CC^*{\cal
P}\theta^2\bar{\theta}^2\,\right)\right. \nonumber \\
&&\left. \qquad+\left(\int d^2\theta \left(\,
\frac{1}{4}f_aW^{a\alpha}W^a_\alpha+C^3W\,\right)+{\rm
h.c.}\right)\, \right]. \eea

Although written in the rigid superspace, the  action
(\ref{rigid-action}) includes all SUGRA effects on SUSY breaking.
Also as it has been derived from locally supersymmetric action
without any inconsistent truncation, it provides a fully
consistent low energy description of KKLT compactification which
contains a red-shifted anti-brane.
  It is obvious
that the uplifting anti-brane operator ${\cal
P}\theta^2\bar{\theta}^2$ does not modify the solutions for the
auxiliary components of $N=1$ superfields, thus \bea
\frac{F^C}{C_0} &=& \frac{1}{3}F^I\partial_IK
+ \frac{C_0^{*2}}{C_0}\, e^{K/3}W^*
\,=\, \frac{1}{3}F^I\partial_IK+m_{3/2}^*, \nonumber \\
F^I&=& -\frac{C^{*2}_0}{C_0}\, e^{K/3}K^{I\bar{J}}(D_J W)^*
\,=\,-e^{K/2}K^{I\bar{J}}(D_JW)^*, \nonumber \\
D^a &=& -C_0C_0^* e^{-K/3}\eta_a^I\partial_IK
\,=\,-\eta^I_a\partial_IK, \eea where $m_{3/2}=e^{K/2}W$ and we
have chosen the Einstein frame condition $C_0=e^{K/6}$ for the
last expressions. Here the index $I$ stands for generic chiral
superfield $\Phi^I$, and  $\eta^I_a$ is the holomorphic Killing
vector for the infinitesimal gauge transformation: \bea
\delta_a\Phi^I=i\alpha_a(x)\eta_a^I. \eea Although it does not
modify the on-shell expression of the auxiliary components of the
$N=1$ superfields, the uplifting operator provides an additional
scalar potential $V_{\rm lift}$ which plays the role of an
uplifting potential in KKLT compactification: \bea V_{\rm TOT}
=V_F+V_D+V_{\rm lift}, \eea where
\begin{eqnarray}
V_F &=& (C_0C_0^*)^2\, e^{K/3} \left\{K^{I\bar{J}}(D_I W)(D_J W)^*
-3|W|^2\right\} \nonumber \\
&=& e^K\left\{K^{I\bar{J}}(D_I W)(D_J W)^* -3|W|^2\right\},
\nonumber \\
V_D &=& \frac{1}{2{\rm Re}(f_a)} D^a D^a, \nonumber \\ V_{\rm
lift} &=& (C_0C_0^*)^2\, {\cal P}\,=\, e^{2K/3}{\cal P}, \eea
where again the last expressions correspond to
the results in the Einstein frame with $C_0=e^{K/6}$.

Let us now consider the KKLT stabilization of CY moduli $\Phi$ and
the resulting soft SUSY breaking terms of visible fields using the
4D effective action (\ref{superaction}) or equivalently
(\ref{rigid-action}). In the first stage, $\Phi$ is stabilized at
the SUSY AdS minimum $\Phi_0$ of $V_{N=1}=V_F+V_D$ for which \bea
D_I W(\Phi_0)=0, \quad W(\Phi_0)\neq 0. \eea The moduli masses at
this SUSY AdS vacuum are dominated by the supersymmetric
contribution which is presumed to be significantly larger than the
gravitino mass: \bea |m_\Phi|\gg |m_{3/2}|,\eea where
 \bea \label{modulimass} m_{\Phi} \simeq
-\left(\frac{e^{K/2}\partial_\Phi^2W}{\partial_\Phi\partial_{\bar{\Phi}}K}\right)_{\Phi_0}.\eea
Adding the uplifting potential will shift the moduli vacuum values
while making the total vacuum energy density to be nearly zero.
Expanding the effective lagrangian of $\Phi$ around $\Phi_0$, one
finds \bea {\cal
L}_\Phi&=&\partial_\Phi\partial_{\bar{\Phi}}K(\Phi_0)\Big(
|\partial_\mu\Delta\Phi|^2-|m_\Phi|^2|\Delta\Phi|^2\Big)
+3|m_{3/2}|^2M_{Pl}^2
\nonumber \\
&&-\,V_{\rm lift}(\Phi_0)-\Big(\Delta\Phi\partial_\Phi V_{\rm lift}(\Phi_0)+{\rm h.c}\Big)
+...,
\eea
where $\Delta\Phi=\Phi-\Phi_0$.
 Then the moduli vacuum shift  is determined to be \bea
\label{shift} \frac{\Delta\Phi}{\Phi_0}\simeq
-\frac{\Phi^*_0\partial_{\bar{\Phi}}V_{\rm
lift}(\Phi_0)}{|\Phi_0|^2\partial_\Phi\partial_{\bar{\Phi}}K(\Phi_0)|m_\Phi|^2}={\cal
O}\left(\frac{m_{3/2}^2}{m_\Phi^2}\right)\eea for
$|\Phi_0|^2\partial_\Phi\partial_{\bar{\Phi}}K(\Phi_0)={\cal
O}(1)$ and $\Phi^0\partial_\Phi V_{\rm lift}(\Phi_0)= {\cal
O}(V_{\rm lift}(\Phi_0))={\cal O}(m_{3/2}^2M_{Pl}^2)$. This vacuum
shift induces a nonzero $F^\Phi$ as \bea \label{fcomponent}
\frac{F^\Phi}{\Phi}&\simeq& \frac{\Delta\Phi\partial_\Phi
F^\Phi+\Delta\Phi^*\partial_{\bar{\Phi}}F^\Phi}{\Phi} \nonumber
\\
&\simeq& -\frac{e^{K/2}(\partial^2_\Phi
W)^*}{\partial_\Phi\partial_{\bar{\Phi}}K}
\frac{\Delta\Phi^*}{\Phi}
\nonumber \\
&\simeq& -\left(\frac{3\Phi^*\partial_\Phi\ln(V_{\rm
lift})}{|\Phi|^2\partial_\Phi\partial_{\bar{\Phi}} K}
\right)\frac{|m_{3/2}|^2}{m_\Phi} \,=\,{\cal
O}\left(\frac{m_{3/2}^2}{m_\Phi}\right), \eea where we have used
$V_{\rm lift}(\Phi_0)\simeq 3|m_{3/2}|^2M_{Pl}^2$. The above
result implies that heavy CY moduli with $m_\Phi\gg 8\pi^2m_{3/2}$
are not relevant for the low energy SUSY breaking since the
corresponding $F^\Phi/\Phi$ is negligible even compared to the
anomaly mediated soft masses of ${\cal O}(m_{3/2}/8\pi^2)$.

 In KKLT compactification, all complex
structure moduli (and the Type IIB dilaton also) are assumed to
get a heavy mass of ${\cal O}(M_{KK}^3/M_{\rm st}^2)$ which is
much heavier than $8\pi^2m_{3/2}$ for $m_{3/2}\lesssim {\cal
O}(8\pi^2)$ TeV.  (Here $M_{KK}$ and $M_{\rm st}$ are the CY
compactification scale and the string scale, respectively.) This
means that complex structure moduli (and the Type IIB dilaton) are
{\it not} relevant for the low energy soft terms, thus can be
safely integrated out. On the other hand, the K\"ahler moduli
masses from hidden gaugino condensations are given by $m_\Phi\sim
m_{3/2}\ln(M^2_{Pl}/m^2_{3/2})$, and thus $F^\Phi/\Phi\sim
m_{3/2}/8\pi^2$. As a result, the K\"ahler moduli can be an
important  messenger of SUSY breaking and generically their
contributions to soft terms are comparable to the anomaly
mediation \cite{choi1}.

The eqs.(\ref{shift}) and (\ref{fcomponent}) show that one needs
to know how the uplifting operator ${\cal P}$ depends on $\Phi$ in
order to determine $F^\Phi/\Phi$. The above discussion implies
also that only the dependence of ${\cal P}$ on the relatively
light moduli with $m_\Phi\lesssim {\cal O}(8\pi^2m_{3/2})$ is
relevant for the low energy SUSY breaking. In KKLT
compactification of Type IIB string theory, anti-brane is
stabilized at the end of a nearly collapsing 3-cycle. On the other
hand, the messenger K\"ahler moduli correspond to the 4-cycle
volumes, thus their wavefunctions have a negligible value at the
end of the collapsing 3-cycle. This implies that K\"ahler moduli
$\Phi$ are {\it sequestered} from anti-brane, i.e. $\partial_\Phi
\ln{\cal P}\simeq 0$. Indeed, in this case, one finds
$K=-3\ln(T+T^*)$ and $V_{\rm lift}\propto 1/(T+T^*)^2$ for the CY
volume modulus $T$, for which ${\cal P}=e^{-2K/3}V_{\rm lift}$ is
independent of $T$.
On the other hand, in the absence of warped throat, one finds
$V_{\rm lift}\propto 1/(T+T^*)^3$ and thus ${\cal P}\propto
1/(T+T^*)$, showing that $T$ has a contact interaction with
anti-brane. This indicates that the presence of warped throat  is
crucial for the sequestering  as well as for the necessary
red-shift of anti-brane. In this paper, although we are mainly
interested in the sequestered anti-brane, we will leave it open
possibility that ${\cal P}$ depends on some messenger moduli.

 To derive the expression of the soft SUSY breaking terms of
visible fields, let us expand $K$ and $W$ in powers of the visible
chiral matter fields $Q^i$: \bea K&=& {\cal
K}_0(\Phi^x,\Phi^{x*},V) +
Z_i(\Phi^x,\Phi^{x*},V)Q^{i*}e^{2q_iV}Q^i, \nonumber \\
W&=&W_0(\Phi^x)+\frac{1}{6}\tilde{\lambda}_{ijk}(\Phi^x)Q^iQ^jQ^k,
\eea where $\Phi^x$ stand for generic messenger superfields of
SUSY breaking, and $V$ is the vector superfield for gauge field.
The soft SUSY breaking terms of canonically normalized visible
fields can be written as
\begin{eqnarray}
{\cal L}_{\rm
soft}&=&-\frac{1}{2}M_a\lambda^a\lambda^a-\frac{1}{2}m_i^2|\tilde{Q}^i|^2
-\frac{1}{6}A_{ijk}y_{ijk}\tilde{Q}^i\tilde{Q}^j\tilde{Q}^k+{\rm
h.c.},
\end{eqnarray}
where $\lambda^a$ are gauginos, $\tilde{Q}^i$ is the scalar
component of the superfield $Q^i$, and $y_{ijk}$ are the
canonically normalized Yukawa couplings:
\begin{eqnarray}
y_{ijk}=\frac{\tilde{\lambda}_{ijk}}{\sqrt{e^{-{\cal
K}_0}Z_iZ_jZ_k}}.
\end{eqnarray}
Then from the superspace action (\ref{rigid-action}), one finds
that the soft masses renormalized at just below the GUT threshold
scale $M_{GUT}$ are given by\footnote{Note that these soft terms are
the consequence of either a non-renormalizable interaction suppressed by
$1/M_{Pl}$ or an exchange of messenger field with a mass close to $M_{Pl}$.
As a result, the messenger scale of these soft terms is close to $M_{Pl}$
although the cutoff scale of the dynamical origin of SUSY breaking,
i.e. the anti-brane,
is $e^AM_{Pl}\sim \sqrt{m_{3/2}M_{Pl}}$ which is far below $M_{Pl}$.}
\begin{eqnarray}
\label{soft1} M_a&=& F^x\partial_x\ln\left({\rm
Re}(f_a)\right) +\frac{b_ag_a^2}{8\pi^2}\frac{F^C}{C_0},
\nonumber \\
A_{ijk}&=&
-F^x\partial_x\ln\left(\frac{\tilde{\lambda}_{ijk}}{e^{-{\cal
K}_0}Z_iZ_jZ_k}\right)-
\frac{1}{16\pi^2}(\gamma_i+\gamma_j+\gamma_k)\frac{F^C}{C_0},
\nonumber \\
m_i^2&=& \frac{2}{3}\langle V_F+V_{\rm lift}\rangle -F^x
F^{x*}\partial_x\partial_{\bar{x}}\ln \left(e^{-{\cal
K}_0/3}Z_i\right)
- \Big(q_i+\eta^x\partial_x\ln(Z_i)\Big)g^2\langle
D\rangle\nonumber
\\&&-\,\frac{1}{32\pi^2}\frac{d\gamma_i}{d\ln\mu}\left|\frac{F^C}{C_0}\right|^2
 + \frac{1}{16\pi^2}\left\{ (\partial_{x}{\gamma}_i)
F^x\left(\frac{F^C}{C_0}\right)^* +{\rm h.c.}\right\}
\nonumber \\
&=&\frac{2}{3}\langle V_{\rm lift}\rangle+\Big( \langle
V_F\rangle+ m_{3/2}^2-F^x F^{x*}\partial_x\partial_{\bar{x}}\ln
\left(Z_i\right)\Big) - \Big(q_i+\eta^x\partial_x\ln(Z_i)\Big)g^2\langle
D\rangle \nonumber
\\&&-\,\frac{1}{32\pi^2}\frac{d\gamma_i}{d\ln\mu}\left|\frac{F^C}{C_0}\right|^2
 + \frac{1}{16\pi^2}\left\{ (\partial_x{\gamma}_i)
F^x\left(\frac{F^C}{C_0}\right)^* +{\rm h.c.}\right\},
\end{eqnarray} where $\partial_x=\partial/\partial\Phi^x$ and $F^x$ is the $F$-component
of $\Phi^x$ which can be determined by (\ref{shift}) and
(\ref{fcomponent}) in KKLT moduli stabilization scenario. Here we
have included the anomaly mediated contributions, i.e. the parts
involving $F^C$, and the $D$-term contribution (for $U(1)$ gauge
group under which $\delta\Phi^x=i\alpha\,\eta^x$) as well as the contributions from $F^x$. As we will see in
the next sections, all these three contributions can be comparable
to each other in models with anomalous $U(1)_A$, thus should be
kept altogether. Here $b_a$ and $\gamma_i$ are the one-loop beta
function coefficients and the anomalous dimension of $Q^i$,
respectively, defined by \bea \frac{dg_a}{d\ln
\mu}=\frac{b_a}{8\pi^2} g_a^3, \qquad \frac{d\ln Z_i}{d\ln
\mu}=\frac{1}{8\pi^2}\gamma_i. \nonumber \eea More explicitly,
\bea b_a&=&-\frac{3}{2}{\rm tr}\left(T_a^2({\rm
Adj})\right)+\frac{1}{2}\sum_i {\rm tr}\left(T^2_a(Q^i)\right),
\nonumber \\
\gamma_i&=&2C_2(Q^i)-\frac{1}{2}\sum_{jk}|y_{ijk}|^2 \quad
(\,\sum_a g_a^2T_a^2(Q^i)\equiv C_2(Q^i)\bf{1}\,),
\nonumber \\
\partial_x\gamma_i
&=&-\frac{1}{2}\sum_{jk}|y_{ijk}|^2\partial_x\ln\left(
\frac{\tilde{\lambda}_{ijk}}{e^{-{\cal K}_0}Z_iZ_jZ_k}\right)
-2C_2(Q^i)\partial_x\ln\left({\rm Re}(f_a)\right), \eea where
$\omega_{ij}=\sum_{kl}y_{ikl}y^*_{jkl}$ is assumed to be diagonal.
Note that  soft scalar masses depend on $\langle V_{\rm
lift}\rangle$, $\langle V_F\rangle$ and $\langle D\rangle$. Since
any of $\langle V_F\rangle$, $\langle V_{\rm lift}\rangle$ and
$\langle D\rangle$ can give an important contribution to $m_i^2$
under the condition of vanishing cosmological constant: \bea
\langle V_{\rm TOT}\rangle =\langle V_F\rangle +\langle V_{\rm
lift}\rangle +\langle V_D\rangle=0,\eea all of these contributions
should be included with correct coefficients.

\section{Mass scales, $F$ and $D$ terms in 4D SUGRA with anomalous $U(1)$}

In this section, we discuss the mass scales and SUSY breaking $F$
and $D$ terms in 4D effective SUGRA which has an anomalous
$U(1)_A$ gauge symmetry. To apply our results to the KKLT
stabilization of the GS modulus, we will include the uplifting Goldstino superfield
operator ${\cal P}\Lambda^2\bar{\Lambda}^2$ which was discussed in
the previous section. The results for the conventional 4D SUGRA
can be obtained by simply taking
the limit ${\cal P}=0$.

In addition to the visible matter
superfields $\{Q^i\}$, the model contains the MSSM singlet
superfields $\{\Phi^x\}=\{\,T,X^p\,\}$ which can participate in SUSY
breaking and/or $U(1)_A$ breaking,
where $T$ is the GS modulus-axion superfield.
 These chiral superfields transform under $U(1)_A$ as \bea
\label{nonlinear} U(1)_A: \quad \delta_A T=-i\alpha(x)\frac{\delta_{GS}}{2}\,,
\quad \delta_A X^p=i\alpha(x)q_p X^p,
\quad \delta_AQ^i=i\alpha(x)q_iQ^i,\eea where $\alpha(x)$
is the infinitesimal $U(1)_A$ transformation function, and
$\delta_{GS}$ is a constant.
 We will choose the normalization of $T$
for which the holomorphic gauge kinetic functions are given by
\bea f_a=k_aT +T\mbox{-independent part},\eea where $k_a$ are real
(quantized) constants of order unity. Under this normalization, we
need $|\langle T\rangle|\lesssim {\cal O}(1)$ to get the gauge
coupling constants of order unity, and also the cancellation of
anomalies by the $U(1)_A$ variation of $k_a{\rm
Im}(T)F^{a\mu\nu}\tilde{F}^a_{\mu\nu}$ requires
\bea
\delta_{GS}={\cal O}\left(\frac{1}{8\pi^2}\right).\eea
Models with anomalous
$U(1)_A$ gauge symmetry contain also an approximate {\it global}
$U(1)$ symmetry: \bea
\label{u1t} U(1)_T: \quad \delta_T T= i\beta,
\quad\delta_T X^p=\delta_TQ^i=0,\eea where $\beta$ is an infinitesimal
constant. Obviously $U(1)_T$ is explicitly broken by
$\delta_Tf_a=ik_a\beta$ as well as by
non-perturbative effects depending on $e^{-cT}$
with an appropriate constant $c$. In some cases, it is more
convenient to consider the following approximate global symmetry
\bea
\label{u1x} U(1)_X:\quad
 \delta_{X}T=0,\quad
\delta_{X}X^p=i\beta q_p X^p,\quad \delta_XQ^i=i\beta
q_iQ^i\eea which is a combination of $U(1)_A$ and $U(1)_T$. The
fact that quantum amplitudes are free from $U(1)_A$ anomaly
requires \bea \label{xanomaly} \left(\delta_X f_a\right)_{\rm
1-loop}= i\beta k_a \frac{\delta_{GS}}{2}\,, \eea
where $\delta_Xf_a$ represent the $U(1)_X$ anomalies due to the fermion loops.

For generic 4D SUGRA action (\ref{superaction}) including
the Goldstino superfield operator ${\cal
P}\Lambda^2\bar{\Lambda}^2$,
one can find the following relation between the vacuum expectation
values of SUSY breaking quantities:
\begin{eqnarray}
\label{relation} && \left( V_F + \frac{2}{3}V_{\rm
lift} + 2|m_{3/2}|^2 + \frac{1}{2}\,M^2_A \right)D_A
\nonumber \\
&& =\, -F^IF^{J*}\partial_I(\eta^L\partial_L\partial_{\bar J}K) +
V_D\eta^I \partial_I\ln g^2_A + V_{\rm lift}\eta^I\partial_I
\ln{\cal P},
\end{eqnarray}
where
$g_A$, $D_A$, $M_A$ and $\eta^I$  denote the gauge coupling,
$D$-term, gauge boson mass, and
holomorphic Killing vector of $U(1)_A$, respectively:
\bea D_A=-\eta^I\partial_IK, \quad
M_A^2=2g_A^2\eta^I\eta^{J*}\partial_I\partial_{\bar{J}}K
\eea
for the $U(1)_A$ transformation
$\delta_A\Phi^I=i\alpha(x)\eta^I$.
Here $V_F$, $V_D$ and $V_{\rm lift}$ are the $F$-term
potential, the $D$-term potential and the uplifting potential,
respectively:
 \bea
V_F&=&K_{I\bar{J}}F^IF^{J*}-3|m_{3/2}|^2,
\nonumber \\
V_D&=&\frac{1}{2}g_A^2D_A^2,
\quad
V_{\rm lift}\,=\,e^{2K/3}{\cal P}, \eea  and all quantities
are evaluated for the vacuum configuration satisfying \bea
\label{stationary}
\partial_IV_{\rm TOT}=
\partial_I(V_F+V_D+V_{\rm lift})=0.
\eea

The  relation (\ref{relation}) has been derived before \cite{kawamura} for
the conventional 4D SUGRA without $V_{\rm lift}$. Since
it plays an important role for our
subsequent discussion, let us briefly sketch the derivation of (\ref{relation}) for
SUGRA including the uplifting
operator ${\cal P}\Lambda^2\bar{\Lambda}^2$. From the $U(1)_A$
invariances of $K$ and $W$, one easily finds \bea
\label{invariance}
\eta^I\partial_IK=\eta^{I*}\partial_{\bar{I}}K,\quad
 \eta^ID_IW=-WD_A\eea
which lead to \bea
&&\eta^I\partial_ID_A=-\eta^I\eta^{J*}\partial_I\partial_{\bar{J}}K=
-\frac{M_A^2}{2g_A^2}\,,
\nonumber \\
&&(\partial_L\eta^I)D_IW+\eta^I\partial_I(D_LW)=W\eta^{\bar{I}}\partial_{\bar{I}}\partial_LK.
\eea Using these relations, one can find
  \bea
\label{gaugeinvariance}
 \eta^I\partial_I V_D&=&
V_D\eta^I\partial_I\ln(g_A^2)-\frac{1}{2}M_A^2D_A, \nonumber \\
\eta^I\partial_IV_F&=&-(V_F+2|m_{3/2}|^2)D_A-F^IF^{J*}\partial_I(\eta^L\partial_L
\partial_{\bar{J}}K),\nonumber \\
 \eta^I\partial_IV_{\rm lift}&=& \left(-
\frac{2}{3}D_A+\eta^I\partial_I\ln ({\cal P})\right)V_{\rm
lift}.\eea Applying the stationary condition
(\ref{stationary}) to (\ref{gaugeinvariance}), one finally obtains the
relation (\ref{relation}).

For the analysis of SUSY and $U(1)_A$ breaking, we can simply set
$Q^i=0$.
 Also for simplicity, we assume that all moduli other than $T$
can be integrated out without affecting  the SUSY and $U(1)_A$
breaking. Then $X^p$ correspond to the $U(1)_A$ charged but MSSM
singlet chiral superfields with vacuum expectation values
 which are small enough to allow the expansion in
powers of $X^p/M_{Pl}$, but still large enough to play an
important role in SUSY and/or $U(1)_A$ breaking. To be concrete,
we will use the K\"ahler potential which takes the form: \bea
\label{kahleru1} K&=&{\cal K}_0(\Phi^x,\Phi^{x*},V_A)+
Z_i(t_V)Q^{i*}e^{2q_iV_A}Q^i\nonumber \\
&=&
K_0(t_V)+Z_p(t_V)X^{p*}e^{2q_pV_A}X^p+Z_i(t_V)Q^{i*}e^{2q_iV_A}Q^i,
\eea where $t_V=T+T^*-\delta_{GS}V_A$ for the $U(1)_A$ vector
superfield $V_A$, however our results will be valid for more
general $K$ including the terms higher order in $X^p/M_{Pl}$. For
the above K\"ahler potential, the $U(1)_A$ $D$-term and gauge
boson mass-square are given by \bea D_A&=& \xi_{FI}-
q_p\tilde{Z}_p|X^p|^2,
\nonumber \\
\frac{M_A^2}{2g_A^2}&=&
M_{GS}^2+\left(q_p^2\tilde{Z}_p-\frac{\delta_{GS}}{2}\,
q_p\partial_T\tilde{Z}_p\right)|X^p|^2,
\eea
where $\xi_{FI}$ and $M_{GS}^2$ are the FI $D$-term and
the GS axion contribution to $M_A^2$, respectively: \bea
\xi_{FI}&=& \frac{\delta_{GS}}{2}\,\partial_TK_0, \nonumber \\
M_{GS}^2&=& \frac{\delta_{GS}^2}{4}\,\partial_T\partial_{\bar{T}}K_0, \eea
and \bea q_p\tilde{Z}_p=q_pZ_p-\frac{\delta_{GS}}{2}\,\partial_TZ_p.\eea
If $|\langle T\rangle|\lesssim {\cal O}(1)$ as required for the gauge coupling constants
to be of order unity, the
K\"ahler metric of $T$ is generically of order unity, and then \bea
M_{GS}\sim \delta_{GS}M_{Pl}\sim
\frac{M_{Pl}}{8\pi^2}.\eea On the other hand, the size of
$\xi_{FI}$ depends on the more detailed property of $T$. If ${\rm Re}(T)$ is a dilaton or
a K\"ahler modulus
stabilized at $\langle {\rm Re}(T)\rangle={\cal O}(1)$, we have
$|\xi_{GS}|\simeq M_{GS}^2(T+T^*)/|\delta_{GS}|\sim 8\pi^2
M_{GS}^2$.
In another case that $T$ is a blowing-up modulus of
orbifold singularity stabilized at near the orbifold limit,
the resulting $|\xi_{FI}|\ll M_{GS}^2$.


In view of that the gaugino masses receive the anomaly mediated
contribution of ${\cal O}(m_{3/2}/8\pi^2)$, one needs  $m_{3/2}$
hierarchically lower than $M_{Pl}$, e.g. $m_{3/2}\lesssim {\cal O}(8\pi^2)$ TeV,
 in order to realize the supersymmetric
extension of the standard model at the TeV scale. Since the $U(1)_A$ gauge boson mass
is always rather close to $M_{Pl}$:  \bea
M_A\gtrsim \sqrt{2}g_AM_{GS}\sim \frac{M_{Pl}}{8\pi^2},\eea
let us focus on models with
\bea \label{condition} m_{3/2}\ll M_A,\quad
\langle V_{\rm TOT}\rangle\simeq 0,\eea and examine the  mass scales in such
models. The condition of nearly vanishing cosmological constant  requires that \bea
K_{I\bar{J}}F^IF^{J*}\lesssim {\cal
O}(m_{3/2}^2M_{Pl}^2), \quad V_{\rm lift}\lesssim {\cal
O}(m_{3/2}^2M_{Pl}^2), \eea and then the relation (\ref{relation})
implies \bea
\label{Dbound} |D_A|\,\lesssim\, {\cal O}\left(
\frac{m_{3/2}^2M_{Pl}^2}{M_A^2}\right)\,\lesssim\, {\cal O}((8\pi^2)^2m_{3/2}^2). \eea

It has been pointed out that one might not need to introduce anti-brane
to obtain a dS vacuum if the $D$-term potential $V_D=\frac{1}{2}g_A^2D_A^2$ can compensate
 the negative vacuum energy density $-3m_{3/2}^2M_{Pl}^2$ in $V_F$ \cite{BKQ}.
The second relation of (\ref{invariance}) indicates that $F^I\neq 0$
is required for $D_A\neq 0$,
thus the $D$-term uplifting scenario can {\it not} be
realized for the supersymmetric AdS solution of $V_F$.
However for a SUSY-breaking solution with $F^I\neq 0$,
$V_D$ might play the role of an uplifting potential
making $\langle V_F+V_D\rangle\geq 0$.
The above bound on $D_A$  imposes a severe limitation on such possibility as it
implies that
$V_D$ can {\it not} be an uplifting potential
in SUSY breaking scenarios with $m_{3/2}\ll M_A^2/M_{Pl}$.
In other words,
SUSY breaking models in which $V_D$ plays the role of
an uplifting potential generically predict a rather large
$m_{3/2}\gtrsim {\cal O}(M_A^2/M_{Pl})\gtrsim {\cal O}(M_{Pl}/(8\pi^2)^2)$.
For instance, the model of \cite{casas} in which $V_D$ indeed compensates
$-3m_{3/2}^2M_{Pl}^2$ in $V_F$
gives $M_A= {\cal O}(M_{Pl}/\sqrt{8\pi^2})$
and $m_{3/2}={\cal O}(M_{Pl}/8\pi^2)$.

Let us now examine more detailed relations between the $F$ and $D$
terms for the K\"ahler potential (\ref{kahleru1}). In case that
$\langle {\rm Re}(T)\rangle={\cal O}(1)$,  the FI $D$-term is rather
close to $M_{Pl}^2$: $\xi_{FI}=
{\cal O}(M_{Pl}^2/8\pi^2)$. Such a large value of $\xi_{FI}$ in $D_A$ should
be cancelled by $q_p\tilde{Z}_p|X^p|^2$ in order to give $D_A$
satisfying the bound (\ref{Dbound}), thus \bea
\label{cancellation} \xi_{FI}\simeq q_p\tilde{Z}_p|X^p|^2.\eea In
some case, for instance the case that the GS modulus is a blowing-up
mode of orbifold singularity,  $\xi_{FI}$ can have a vacuum value
smaller than $M_{Pl}^2$ by many orders of magnitude. However the
existence of the anomalous (approximate) global symmetry $U(1)_X$
implies that some $X^p$ should get a large vacuum value
$|X^p|^2\gg |D_A|$ to break $U(1)_X$ at a sufficiently high energy scale.
This means that $|\xi_{FI}|\gg |D_A|$ and
 the relation
(\ref{cancellation}) remains to be valid
even in case that  $|\xi_{FI}|$ is smaller than $M_{Pl}^2$ by many orders of
magnitude.
Then using $\eta^ID_IW=-WD_A$, we find \bea
\label{frelation} F^T&=&
\frac{q_p\tilde{Z}_p|X^p|^2}{\delta_{GS}\partial_T\partial_{\bar{T}}K_0/2
-q_r\partial_T\tilde{Z}_r|X^r|^2}\left(\frac{F^p}{X^p}\right)+
{\cal O}\left(\frac{8\pi^2m_{3/2}D_A}{M_{Pl}}\right)
\nonumber \\
&=&\frac{{\cal O}(\delta_{GS}\xi_{FI})}{M_{GS}^2+{\cal
 O}(\delta_{GS}\xi_{FI})}\frac{F^p}{X^p},\eea
where we have used (\ref{cancellation}) for the last expression.
Applying this relation to (\ref{relation}), we also find \bea
\label{drelation} g_A^2D_A&=&
-\frac{q_p\tilde{Z}_p\delta_{p\bar{q}}+q_pq_qX^{p*}X^q\partial_T[
\tilde{Z}_p\tilde{Z}_q/(\delta_{GS}\partial_T\partial_{\bar{T}}K_0/2
-q_r\partial_T\tilde{Z}_r|X^r|^2)]}
{\delta_{GS}^2\partial_T\partial_{\bar{T}}K_0/4+(q^2_r\tilde{Z}_r-q_r\delta_{GS}\partial_T\tilde{Z}_r/2)
|X^r|^2}{F^p}{F^{q*}} \nonumber \\
&& +\,\frac{V_{\rm lift}\eta^I\partial_I\ln{\cal
P}}{\delta_{GS}^2\partial_T\partial_{\bar{T}}K_0/4+
(q^2_r\tilde{Z}_r-q_r\delta_{GS}\partial_T\tilde{Z}_r/2)
|X^r|^2}
 \nonumber \\
&=&\frac{{\cal O}(\xi_{FI})}{M_{GS}^2+{\cal
O}(\xi_{FI})}\left|\frac{F^p}{X^p}\right|^2
+\frac{V_{\rm lift}}{M_{GS}^2+{\cal
O}(\xi_{FI})}\eta^I\partial_I\ln{\cal P}. \eea
Note that the piece proportional to $V_{\rm lift}$ vanishes if the Goldstino superfield on anti-brane
is sequestered from the $U(1)_A$ charged fields,
i.e. $\eta^I\partial_I{\cal P}=0$,
which is a rather plausible possibility in view of our discussion in section 2.

The relations (\ref{frelation}) and (\ref{drelation}) show that the relative importance
of the GS modulus mediation and the $U(1)_A$ $D$-term mediation
is determined essentially by the ratio
\bea
R\equiv \frac{\xi_{FI}}{M_{GS}^2}=\frac{2\partial_TK_0}{\delta_{GS}
\partial_T\partial_{\bar{T}}K_0}.
\eea
If $T$ is a string dilaton or a K\"ahler modulus  normalized as $\partial_Tf_a={\cal O}(1)$, its
K\"ahler potential is given by
$$K_0=-n_0\ln(T+T^*)+{\cal O}(1/8\pi^2(T+T^*)).$$
As long as ${\rm Re}(T)$ is stabilized at a value of ${\cal O}(1)$,
the higher order string loop or $\alpha^\prime$
corrections to $K_0$ can be safely ignored,
yielding
$|R|={\cal O}(8\pi^2)$.
In such case, (\ref{frelation}) and (\ref{drelation}) imply that generically
\bea
|D_A|\sim |F^T|^2\sim \left|\frac{F^p}{X^p}\right|^2.
\eea
Note that in the limit $|R|\gg 1$, the $U(1)_A$ gauge boson mass-square
is dominated by the contribution from $\langle X^p\rangle\sim \sqrt{|\xi_{FI}|}$.
In this case, the longitudinal component of the massive $U(1)_A$ gauge boson
comes mostly from the phase degrees  of $X^p$, while
the GS modulus $T$ is approximately a flat direction of the
$U(1)_A$ $D$-term potential.
An interesting possibility is then to stabilize $T$ by non-perturbative superpotential
at a SUSY AdS vacuum with ${\rm Re}(T)={\cal O}(1)$, and then lift this
AdS vacuum to dS state by adding a red-shifted anti-brane as in the KKLT moduli
stabilization scenario.
In the next section, we will discuss such KKLT stabilization of the GS modulus
in more detail together with
the resulting pattern of soft SUSY breaking terms.

Another  possibility is that ${\rm Re}(T)$ is a blowing-up modulus of orbifold singularity,
thus $\xi_{FI}=\delta_{GS}\partial_TK_0=0$ in the orbifold limit.
Choosing ${\rm Re}(T)=0$ in the orbifold limit, $K_0$ can be expanded as
$$
K_0\approx \frac{1}{2}a_0(T+T^*)^2+{\cal O}((T+T^*)^3)
$$
for a constant $a_0$.
If ${\rm Re}(T)$ is stabilized at near the orbifold limit for which
$|\xi_{FI}| \ll
M_{GS}^2$,
the resulting
$|R|\ll 1$.
In this limit, if the uplifting anti-brane  is sequestered
from the $U(1)_A$ charged fields, i.e.  $\eta^I\partial_I{\cal
P}=0$, eqs. (\ref{frelation}) and (\ref{drelation}) lead to \bea
F^T\,\sim\, \delta_{GS}R\,\frac{F^p}{X^p}, \quad
D_A\,\sim\, R\left|\frac{F^p}{X^p}\right|^2, \eea where
$F^p/X^p$
represents the SUSY breaking mediated at the scale around
$\langle X^p\rangle\sim \sqrt{|\xi_{FI}|}\ll M_{GS}$. The anomaly
condition (\ref{xanomaly}) for the $U(1)_X$
symmetry (\ref{u1x})  implies that the gauge kinetic functions
receive a loop correction $\Delta f_a\sim \frac{1}{8\pi^2}\ln X^p$
at the scale $\langle X^p\rangle$ where $U(1)_X$ is spontaneously
broken. For instance, there might be a coupling  $X^pQ_1Q_2$ in
the superpotential generating $\Delta f_a$ through the loop of
$Q_1+Q_2$ which are charged under the standard model gauge
group.\footnote{This corresponds to the gauge mediation at the
messenger scale $\langle X^p\rangle$.}
 This results in the gaugino masses \bea M_a ={\cal
O}\left(\frac{1}{8\pi^2}\frac{F^p}{X^p}\right) \eea mediated at
the scale  $\langle X^p\rangle$. Obviously $F^T$ is smaller than
this $M_a$ in the limit $|R|\ll 1$.
If $\xi_{FI}$ is smaller than $M_{Pl}^2$ by many orders of
magnitude, e.g. $|R|\lesssim 10^{-4}$, $|D_A|$ also is smaller
than $|M_a|^2$ mediated at $\langle X^p\rangle$. Then the soft
terms are dominated by the contributions mediated at the low
messenger scale around $\langle X^p\rangle\sim \sqrt{|\xi_{FI}|}$.
Those soft terms with low messenger scale depend on more detailed
property of the model, which is beyond the scope of this paper.

\section{A model for the KKLT stabilization of the GS modulus}

In this section, we discuss a model
for the KKLT stabilization of the GS modulus $T$
in detail.
In this model, $T$ is stabilized at a value  of ${\cal O}(1)$,
yielding $\xi_{FI}\sim \delta_{GS}M_{Pl}^2$.
For simplicity, we introduce a single $U(1)_A$ charged
MSSM singlet $X$ whose
vacuum value cancels $\xi_{FI}$ in $D_A$.
In addition to $X$ and the visible matter superfields $Q^i$,
one needs also a hidden $SU(N_c)$ Yang-Mills sector with $SU(N_c)$
charged  matter fields $Q_H+Q^c_H$ in order to
produce non-perturbative superpotential stabilizing $T$.
The gauge kinetic functions of the model are given by
\bea
\label{gaugekinetic}
f_a=kT+\Delta f, \qquad f_H=k_HT+\Delta f_{H},
\eea
where $f_a$ ($a=3,2,1$) and $f_H$ are the gauge kinetic functions of the $SU(3)_c\times SU(2)_W\times
U(1)_Y$ and the hidden $SU(N_c)$ gauge group, respectively, and
$k$ and $k_H$ are real constants of ${\cal O}(1)$.
Generically $\Delta f$ and $\Delta f_{H}$ can depend on other moduli of the model.
Here we assume that those other moduli are fixed by fluxes with a large mass
$\gg 8\pi^2 m_{3/2}$, and then  $\Delta f$ and $\Delta f_{H}$ can be considered as constants
which are obtained by integrating out the heavy moduli.

The K\"ahler potential, superpotential,
and the uplifting operator are given by
\begin{eqnarray}
\label{model} && K \,=\, K_0(t_V) + Z_X(t_V)X^*e^{-2V_A}X +
Z_H(t_V)Q_H^*e^{2qV_A}Q_H
\nonumber \\
&& \qquad \,\,+\,Z^c_H(t_V)Q^{c*}_He^{2q_{c}V_A}Q^c_H
+Z_{i}(t_V)Q^{i*}e^{2q_iV_A}Q^i, \nonumber \\
&& W \,=\, \omega_0 + \lambda X^{q+q_c}Q_H^cQ_H +
(N_c-N_f) \left(\frac{e^{-8\pi^2 f_H}}{{\rm
det}(Q_H^cQ_H)}\right)^{\frac{1}{N_c-N_f}} \nonumber \\
&&\qquad\,\, +\,
\frac{1}{6}\,\lambda_{ijk}X^{q_i+q_j+q_k}Q^i Q^j Q^k, \nonumber \\
&&
{\cal P}\,=\,{\cal P}(t_V),
\end{eqnarray}
where $t_V=T+T^*-\delta_{GS}V_A$, $w_0$ is a constant of ${\cal
O}(m_{3/2}M_{Pl}^2)$, $\lambda$ and $\lambda_{ijk}$ are constant
Yukawa couplings, $N_f$ denotes the number of flavors for the
hidden matter $Q_H+Q^c_H$, ${\cal P}\Lambda^2\bar{\Lambda}^2$ is
the uplifting Goldstino superfield operator induced by anti-brane,
and finally the $U(1)_A$ charge of $X$ is normalized as $q_X=-1$.
As we have discussed in section 2, anti-brane in KKLT
compactification is expected to be sequestered from the $D$-brane
of $U(1)_A$, and then ${\cal P}$ is independent of $t_V$. Here we
consider more general case that ${\cal P}$ can depend on $t_V$ in
order to see what would be the consequence of the uplifting
operator if it is not sequestered from $U(1)_A$. Note that the GS
cancellation of the mixed anomalies of $U(1)_A$ requires \bea
\label{gscondition}
\frac{\delta_{GS}}{2} =\frac{N_f(q+q^c)}{8\pi^2k_H}=\frac{\sum_iq_i{\rm
Tr}(T_a^2(Q^i))}{4\pi^2k_a}. \eea

In our case, the non-perturbative superpotential in (\ref{model}),
i.e. the Affleck-Dine-Seiberg superpotential \cite{ADS}\bea W_{\rm
ADS}=(N_c-N_f) \left(\frac{e^{-8\pi^2 f_H}}{{\rm
det}(Q_H^cQ_H)}\right)^{\frac{1}{N_c-N_f}} \eea
 requires a more careful interpretation.
If $\lambda$ is so small that the tree level mass
$M_Q=\lambda\langle X^{q+q_c}\rangle$
 of $Q_H+Q^c_H$
 is lower than the dynamical scale  of
$SU(N_c)$ gauge interaction, $W_{\rm ADS}$ can be interpreted as
the non-perturbative superpotential of the  light composite meson
superfields $\Sigma=Q_H^cQ_H$. However a more plausible possibility
is that $\lambda={\cal O}(1)$, and so (in the unit with
$M_{Pl}=1$) \bea M_Q=\lambda\langle X^{q+q_c}\rangle \gg
\Lambda_H= \Big(e^{-8\pi^2 f_H}{\rm det}(M_Q)\Big)^{1/(3N_c)}. \eea
Note that $|X|^2={\cal O}(|\xi_{FI}|)={\cal O}(M_{Pl}^2/8\pi^2)$
in this model. In this case, the correct procedure to deal with
$SU(N_c)$ dynamics is  to integrate out first the heavy
$Q_H+Q_H^c$ at the scale $M_Q$. The resulting effective theory is
a pure super YM theory at the scale just below $M_Q$, but
with the modified gauge kinetic function: \bea f_{\rm eff}(M_Q)=
f_H+\frac{3N_c-N_f}{8\pi^2}\ln(M_Q/M_{Pl}). \eea
 Then the
$SU(N_c)$ gaugino condensation is formed at $\Lambda_H$ by this
pure super YM dynamics, yielding a non-perturbative superpotential
\bea W_{\rm eff}=N_c M_Q^3e^{-8\pi^2 f_{\rm eff}(M_\Phi)/N_c}.
\eea This $W_{\rm eff}$ is the same as the non-perturbative
superpotential obtained by integrating out  $\Sigma=Q^c_HQ_H$
using the equations of motion $\partial_\Sigma W=0$ for the
superpotential of (\ref{model}). In the following, we will simply
use the superpotential of (\ref{model}) since it leads to the
correct vacuum configuration independently of the value of
$M_Q/\Lambda_H$.

To examine the vacuum configuration of the model (\ref{model}),
it is convenient to estimate first the mass scales of
the model. As long as $m_{3/2}$
is hierarchically smaller than $M_{Pl}$,
one easily finds that the following mass patterns are independent
of the details of SUSY breaking. First,
$T$ is stabilized at a vacuum expectation value of ${\cal O}(1)$, and as a result
\bea
R=\frac{\xi_{FI}}{M_{GS}^2}=\frac{2\partial_TK_0}{\delta_{GS}
\partial_T\partial_{\bar{T}}K_0}={\cal O}(8\pi^2).
\eea
The $U(1)_A$ gauge boson mass-square is dominated
by the contribution from $|X|^2 \sim |\xi_{FI}|$:
\bea
\frac{M_A^2}{2g^2_A} \simeq Z_X|X|^2+ M_{GS}^2\simeq  Z_X|X|^2.
\eea
The hidden $SU(N_C)$ confines at the scale \bea \Lambda_H =
\Big(e^{-8\pi^2 (k_HT+\Delta f_{H})}{\rm
det}(M_Q/M_{Pl})\Big)^{1/(3N_c)}M_{Pl},\eea and the $SU(N_c)$
$D$-flat directions of the hidden matter fields are stabilized at
\bea \langle Q_H^cQ_H\rangle \sim \frac{\Lambda_H^3}{M_Q}. \eea
Finally the hidden $SU(N_c)$ scale and $m_{3/2}$ obey the
standard relation: \bea \frac{\Lambda_H^3}{M_{Pl}^3}\,\sim\,
\frac{m_{3/2}}{M_{Pl}}. \eea

It is straightforward to see
that in the absence of the uplifting Goldstino operator
${\cal P}\Lambda^2\bar{\Lambda}^2$, the model (\ref{model})
has a unique and stable SUSY AdS vacuum.\footnote{
This SUSY AdS vacuum  is a saddle point solution of
$V_F$, but is the global minimum of $V_F+V_D$.} For $m_{3/2}$
hierarchically smaller than $M_A$ and $\Lambda_H$, adding the
uplifting operator with ${\cal P}\sim
m^2_{3/2}M^2_{Pl}$ triggers a small shift of vacuum configuration,
leading to non-zero vacuum expectation values of $F^T, F^X,
F^\Sigma$ and $D_A$, where $\Sigma=Q_H^cQ_H$.
In the following, we compute
these SUSY breaking vacuum values within a perturbative
expansion in
$$
\frac{\delta_{GS}}{T+T^*} ={\cal
O}\left(\frac{1}{8\pi^2}\right),$$ while ignoring the
corrections suppressed by the following scale hierarchy
factors:\bea \label{scale-hierarchy}
\frac{\Lambda_H}{M_A},\, \frac{m_{3/2}}{\Lambda_H},\,
\frac{m_{3/2}}{M_\Phi}, \frac{\langle Q_H^cQ_H\rangle}{\langle
XX^*\rangle}\, \ll\, \frac{1}{8\pi^2}. \eea

Let us now examine the vacuum configuration in more detail.
As we have mentioned, the true vacuum configuration is given by a small shift
induced by $V_{\rm lift}$ from the
SUSY AdS solution of $D_A=0$ and $D_IW=0$.
With this observation, we find (in the unit with $M_{Pl}=1$):
\bea
\label{vacuumvalue}
|X|^2&=& -\frac{\delta_{GS}\partial_T
K_0}{2Z_X}\left(1+{\cal O}\left(\frac{1}{8\pi^2}\right)\right),
\nonumber \\
Q^c_HQ_H&=& e^{-8\pi^2(k_HT+\Delta f_{H})/N_c}(\lambda X^{q+q_c})^{(N_f-N_c)/N_c},
\nonumber \\
{\rm Re}(T)&= &\frac{N_c}{8\pi^2k_H}\ln
\left|\frac{8\pi^2k_H}{\omega_0\partial_TK_0}\right|
-\frac{\Delta f_{H}+\Delta f_{H}^*}{2k_H}
+\frac{N_f}{8\pi^2k_H}\ln|\lambda X^{q+q_c}|+{\cal O}\left(\frac{1}{8\pi^2}\right)
\nonumber \\
&=& \frac{N_c}{8\pi^2k_H}\ln\left(\frac{M_{Pl}}{m_{3/2}}\right)-\frac{\Delta f_{H}+\Delta f_{H}^*}{2k_H}
+{\cal O}\left(\frac{1}{8\pi^2}\right).
 \eea
Note that $|X|^2={\cal O}(M_{Pl}^2/8\pi^2)$, thus an effect further suppressed by
$|X|^2/M_{Pl}^2$ is comparable to the loop correction.
The above result  on the vacuum expectation value of ${\rm Re}(T)$ shows that
the GS modulus is  stabilized at a value of ${\cal O}(1)$
for the model parameters giving the weak scale SUSY, e.g.
$m_{3/2}\lesssim {\cal O}(8\pi^2)$ TeV.

If ${\rm Re}(T)$ is stabilized at a value of ${\cal O}(1)$ as desired,
$\Sigma=Q^c_HQ_H$ is hierarchically smaller than $M_{Pl}^2$.
Since $F^\Sigma={\cal O}(\Sigma F^T)$
and the couplings between $Q_H+Q^c_H$ and the visible
fields are suppressed by  $1/M_{Pl}$,
the contribution from $F^\Sigma$ to the visible soft terms
can be ignored.
Then  the soft terms of visible fields are determined by
the following four SUSY-breaking auxiliary
components: \bea
\label{aucom}
\frac{F^T}{T+T^*}&=&\frac{m^*_{3/2}}{8\pi^2}\left(
\frac{3N_c\partial_T\ln(V_{\rm lift})}{k_H(T+T^*)\partial_TK_0}\right)
\left(1+{\cal O}\left(\frac{1}{8\pi^2}\right)\right), \nonumber \\
 \frac{F^X}{X}&=&-F^T\partial_T\ln\left(-\frac{Z_X}{\partial_TK_0}\right)
\left(1+{\cal O}\left(\frac{1}{8\pi^2}\right)\right),
\nonumber \\
 g_A^2D_A &=&\left|F^T\right|^2\partial_T\partial_{\bar{T}}\ln\left(-\frac{Z_X}{\partial_TK_0}\right)
\left(1+{\cal O}\left(\frac{1}{8\pi^2}\right)\right)
\nonumber \\
&&+\,V_{\rm lift}\frac{\partial_T\ln {\cal
P}}{\partial_T K_0} \left(1+{\cal O}\left(\frac{1}{8\pi^2}\right)\right),
\nonumber \\
\frac{1}{8\pi^2}\frac{F^C}{C_0}
&=&\frac{m_{3/2}^*}{8\pi^2}\left(1+{\cal
O}\left(\frac{1}{8\pi^2}\right)\right), \eea where  $F^C/8\pi^2$
and $F^T$ are the order parameters of anomaly mediation and GS
modulus mediation, respectively, and $F^X$ and $D_A$ are the order
parameters of $U(1)_A$ mediation. Note that $V_A$ and $X$
constitute a massive vector superfield $\tilde{V}_A=V_A-\ln|X|$.
The results on $F^X$ and $D_A$ can be obtained from eqs.
(\ref{frelation}) and (\ref{drelation}), while the result on $F^T$
can be obtained by applying eqs. (\ref{modulimass}) and
(\ref{fcomponent}).

The above results show that generically the GS modulus mediation,
the anomaly mediation and the $X$ mediation are comparable to each
other. If anti-brane and the $D$-brane of $U(1)_A$ are separated
from  each other by a warped throat, it is expected that
$\partial_T\ln{\cal P}=0$. Then the $U(1)_A$ $D$-term mediation
is also generically comparable to the other mediations. However,
if the K\"ahler potential of $T$ and $X$ has a special form to
give ${Z_X}/{\partial_TK_0}=\mbox{constant}$, we have
${F^X}/{X}={\cal O}(F^T/8\pi^2)$ and $D_A={\cal
O}(|F^T|^2/8\pi^2)$, thus the $U(1)_A$ mediation is suppressed by
a loop factor of ${\cal O}(1/8\pi^2)$ compared to the GS-modulus
and anomaly mediations. Finally, if anti-brane is not sequestered,
the resulting $D_A$ is of ${\cal O}(m_{3/2}^2)={\cal
O}((8\pi^2F^T)^2)$ and then soft sfermion masses are dominated by
the $U(1)_A$ $D$-term contribution. Another important feature of
(\ref{aucom})   is that $F^T$, $F^X/X$ and $F^C/C_0$ are
relatively {\it real} since $K_0, Z_X, {\cal P}$ are real
functions of the real variable $t=T+T^*$. As a result, the gaugino
masses and $A$-parameters mediated by these auxiliary components
automatically preserve CP \cite{choi4}. Since one can always make
$m_{3/2}=e^{K/2}W$ to be real by an appropriate
$R$-transformation, all of the above auxiliary components can be
chosen to be real, which will be taken in the following
discussions.

Applying the above results to the soft terms of (\ref{soft1}) and also taking into account
that $|X|^2/M_{Pl}^2={\cal O}(1/8\pi^2)$, we find the soft masses at
the scale just below $M_{GUT}$:
\bea
\label{kkltsoft}
M_a&=&M_0+\frac{b_a}{8\pi^2}\,g_{GUT}^2m_{3/2}+{\cal O}\left(\frac{M_0}{8\pi^2}\right),
\nonumber \\
A_{ijk}&=&M_0(a_{i}+a_{j}+a_{k})-\frac{1}{16\pi^2}(\gamma_i+\gamma_j+\gamma_k)m_{3/2}
+{\cal O}\left(\frac{M_0}{8\pi^2}\right),
\nonumber \\
m_i^2&=&c_iM_0^2-\frac{1}{32\pi^2}\,\dot{\gamma}_im_{3/2}^2
+\frac{m_{3/2}M_0}{8\pi^2}\left(\frac{1}{2}\sum_{jk}|y_{ijk}|^2(a_i+a_j+a_k)-2C_2(Q^i)\right)
\nonumber \\
&& -\, 3q_im_{3/2}^2\frac{\partial_T\ln{\cal P}}{\partial_TK_0}\left(1+
{\cal O}\left(\frac{1}{8\pi^2}\right)\right)+{\cal O}\left(\frac{M_0^2}{8\pi^2}\right),
\eea
 where $M_0$  is the universal modulus-mediated gaugino mass
at $M_{GUT}$:
\bea
\label{parameter}
M_0\equiv F^T\partial_T\ln {\rm Re}(f_a)= \frac{m_{3/2}}{8\pi^2}\left(\frac{3N_c
\partial_T\ln V_{\rm lift}}{k_H\partial_TK_0}\right)\partial_T\ln({\rm Re}(f_a)),
\eea
for
\bea
\partial_T\ln({\rm Re}(f_a))=\frac{1}{T+T^*+(\Delta f+\Delta f^*)/k}
=\frac{kg_{GUT}^2}{2},
\eea
and
\bea
a_i&=&\frac{\partial_T\ln\left(e^{-K_0/3}Z_i\left(-{Z_X}/{\partial_TK_0}\right)^{q_i}\right)}{
\partial_T\ln({\rm Re}(f_a))},
\nonumber \\
c_i&=&-\frac{\partial_T\partial_{\bar{T}}\ln\left(e^{-K_0/3}Z_i
\left(-{Z_X}/{\partial_TK_0}\right)^{q_i}\right)}{[\partial_T\ln({\rm Re}(f_a))]^2}. \eea

Here $\dot{\gamma_i}=d\gamma_i/d\ln\mu$ for the anomalous dimension
$\gamma_i=8\pi^2d\ln Z_i/d\ln\mu$, $2C_2(Q^i)$ is the gauge
contribution to $\gamma_i$, i.e. $C_2(Q^i){\bf 1}=\sum_a
g_a^2T_a^2(Q^i)$ for the gauge generator $T_a(Q^i)$, and finally
the canonical Yukawa couplings are given by \bea y_{ijk}
=\frac{\lambda_{ijk}(\delta_{GS}/2)^{(q_i+q_j+q_k)/2}}
{\sqrt{(-{Z}_X/\partial_TK_0)^{q_i+q_j+q_k}e^{-K_0}Z_iZ_jZ_k}}.
\eea

The soft parameters of (\ref{kkltsoft}) show that
the gaugino masses $M_a$ in models of KKLT stabilization of the GS modulus
are determined by the GS modulus mediation and the anomaly mediation
which are comparable to each other.
In case that anti-brane is sequestered from $U(1)_A$ and thus from the GS modulus $T$,
i.e. $\partial_T\ln{\cal P}=0$,
soft sfermion masses are comparable to the gaugino masses.
However if anti-brane is not sequestered,
soft sfermion mass-squares (for $q_i\neq 0$) are dominated
by the $U(1)_A$ $D$-term contribution of ${\cal O}(8\pi^2 M_a^2)$, which
might enable us to realize the more minimal supersymmetric standard model scenario \cite{CKN}.

It has been noticed that the low energy gaugino masses obtained
from the renormalization group (RG) running of the gaugino masses
of (\ref{kkltsoft}) at $M_{GUT}$ are given by \bea
M_a(\mu)=M_0\left[1-\frac{1}{4\pi^2}b_ag_a^2(\mu)\ln\left(\frac{M_{GUT}}{(M_{Pl}/m_{3/2})^{\alpha/2}\mu}
\right)\right], \eea which are same as the low energy gaugino
masses in pure modulus-mediation started from the mirage messenger
scale \bea M_{\rm mirage}=(m_{3/2}/M_{Pl})^{\alpha/2}M_{GUT}, \eea
where \bea \alpha\equiv\frac{m_{3/2}}{M_0\ln(M_{Pl}/m_{3/2})}.
\eea Similar mirage mediation pattern arises also for the low
energy soft sfermion masses if the involved Yukawa couplings are
small or $a_i+a_j+a_k=1$ and $c_i+c_j+c_k=1$ for the combination
$(i,j,k)=(H_u,t_L,t_R)$ in the top-quark Yukawa coupling. From
(\ref{parameter}), we find that the anomaly to modulus mediation ratio $\alpha$ is given by
\bea
\label{alpha}
\alpha=\frac{2\partial_TK_0}{2\partial_TK_0+3\partial_T\ln{\cal
P}} \left(\
1+\frac{4\pi^2[k_H(\Delta f+\Delta f^*)-k(\Delta f_{H}+\Delta f_{H}^*)]}{kN_c\ln(M_{Pl}/m_{3/2})}\right).
\eea
In the minimal
KKLT model, anti-brane is sequestered, i.e $\partial_T\ln{\cal P}=0$,
$\Delta f \simeq 0$ and $\Delta f_H\simeq 0$, thus
$\alpha=1$. However in more generic compactifications, it is
possible that the gauge kinetic functions $f_a$ and $f_H$ are
given by different linear combinations of $T$ and other moduli.
Stabilizing the other moduli can give rise to sizable $\Delta f_H$
and/or $\Delta f$,
thus a different value of $\alpha$ even in the case that
anti-brane is sequestered \cite{abe}. In this regard, one
interesting possibility is to have $\alpha=2$ which leads to the TeV scale
mirage mediation. As was noticed in \cite{choi3}, the little
hierarchy problem of the MSSM can be significantly ameliorated in
the TeV scale mirage mediation scenario.
For the model under discussion, $\alpha= 2$ can be achieved for instance
when $\partial_T\ln{\cal P}=0$,
${\rm Re}(\Delta f)=0$ and ${\rm Re}(\Delta f_H)=-N_c/2$.

To be more concrete, let us consider
the following  K\"ahler potential and the uplifting operator which are expected
to be valid for a wide class of string compactifications:
\begin{eqnarray}
f_a&=&kT, \quad f_H\,=\,k_HT+\Delta f_{H},
\nonumber \\
K_0&=& -n_0\ln(t_V), \quad Z_I\,=\, \frac{1}{t_V^{n_I}},\quad
{\cal P}\,=\,\frac{{\cal P}_0}{t_V^{n_P}},
\end{eqnarray}
where $Z_I$ denote the K\"ahler metric of $\Phi^I=(X,Q^i)$, and
${\cal P}_0$ is a constant of ${\cal O}(m_{3/2}^2M_{Pl}^2)$.
Applying (\ref{vacuumvalue}), (\ref{aucom}), (\ref{kkltsoft}) and (\ref{alpha})  to this
form of gauge kinetic functions, K\"ahler
potential and uplifting operator, we find
\begin{eqnarray}
|X|^2
&=& \frac{n_0\delta_{GS}}{2(T+T^*)^{1-n_X}},
\nonumber \\
\frac{F^X}{X} &=& (n_X-1)M_0,
\nonumber \\
g_A^2D_A &=&(n_X-1)M_0^2+ \frac{3n_P}{n_0}m^2_{3/2},
\nonumber \\
a_i&=&c_i\,=\,\frac{1}{3}\,n_0-n_i-(n_X-1)q_i, \nonumber \\
\alpha&=&\frac{2n_0}{2n_0+3n_P}
\left(1-\frac{4\pi^2(\Delta f_{H}+\Delta f_H^*)}
{N_c\ln(M_{Pl}/m_{3/2})}\right),
\nonumber \\
y_{ijk}&=& \frac{(n_0\delta_{GS}/2)^{(q_i+q_j+q_k)/2}}{
(T+T^*)^{(a_i+a_j+a_k)/2}}\lambda_{ijk}.
\end{eqnarray}

In fact, since $U(1)_A$ is spontaneously broken by $\langle X\rangle\sim
M_{Pl}/\sqrt{8\pi^2}$, the soft parameters of (\ref{kkltsoft}) can be
obtained also from an effective SUGRA which would be derived
by integrating out the massive vector
multiplet $\tilde{V}_A=V_A-\ln|X|$ as well as the hidden matter
$Q_H+Q_H^c$. To derive the effective SUGRA, it is convenient to make
the following field redefinition:
\bea
V_A &\rightarrow & V_A+\ln|X|,
\nonumber \\
T &\rightarrow &T+\frac{\delta_{GS}}{2}\ln(X),
\nonumber \\
Q^I &\rightarrow & X^{-q_I}Q^I
\quad (Q^I=Q_H,Q_H^c,Q^i).
\eea
This field redefinition induces an anomalous variation of the gauge kinetic functions
\bea
&&f_a\rightarrow f_a-\frac{1}{4\pi^2}\sum_iq_i{\rm Tr}(T_a^2(Q^i))\ln(X),
\nonumber \\
&&f_H\rightarrow f_H-\frac{1}{8\pi^2}(q+q^c)N_f\ln(X). \eea Taking
into account this change of gauge kinetic functions together with
the anomaly cancellation condition (\ref{gscondition}), the model
in the new field basis is given by \bea \label{remodel}
K &=& K_0(t_V) + Z_X(t_V)e^{-2V_A} +
Z_I(t_V)Q^{I*}e^{2q_IV_A}Q^I
\nonumber \\
W &=& \omega_0 + \lambda Q_H^cQ_H +
\frac{1}{6}\,\lambda_{ijk}Q^i Q^j Q^k, \nonumber \\
f_a&=&kT+\Delta f, \quad f_H=k_HT+\Delta f_{H},\quad {\cal P}\,=\,{\cal
P}(t_V), \eea In the new field basis, $V_A$ corresponds to the
massive vector superfield $\tilde{V}_A$. The heavy hidden matter
$Q_H+Q^c_H$ can be easily integrated out, leaving a threshold
correction to the hidden gauge kinetic function: $\delta
f_H=-N_f\ln(\lambda)/8\pi^2$. The massive vector superfield can be
also integrated out using the equation of motion:
\begin{eqnarray}
\frac{\partial K}{\partial V_A} - \theta^2\bar\theta^2CC^*e^{K/3}
\frac{\partial {\cal P}}{\partial V_A}&=&0.
\end{eqnarray}
For simplicity, here we will consider only the case of sequestered
anti-brane, i.e. $\partial {\cal P}/\partial V_A=0$. The
generalization to unsequestered anti-brane is straightforward.
Making an expansion in $\delta_{GS}={\cal O}(1/8\pi^2)$, the
solution of the above equation is given by \bea
e^{-2V_A}&=&-\frac{\delta_{GS}\partial_TK_0}{2{Z}_X}\left(1+ {\cal
O}\left(\frac{1}{8\pi^2}\right)\right)
\nonumber \\
&&+\, q_i\left(\frac{-2{Z}_X}{\delta_{GS}\partial_TK_0}\right)^{q_i}
\left(1+{\cal O}\left(\frac{1}{8\pi^2}\right)\right)Z_i Q^{i*}Q^i. \eea
Inserting this solution to (\ref{remodel}) and also adding the
gaugino condensation superpotential of the super $SU(N_c)$ YM
theory whose gauge kinetic function is now given by
$f_H=k_HT+\Delta f_{H}+\delta f_H$, we find the following effective
SUGRA: \bea K_{\rm eff}&=&K_0(t)
+\left|\frac{Z_X(t)}{\partial_TK_0(t)}\right|^{q_i}Z_i(t)Q^{i*}Q^i,
\nonumber \\
W_{\rm eff}&=&w_0+N_c\lambda^{N_f/N_c}e^{-8\pi^2\Delta f_{H}/N_c}e^{
-8\pi^2k_HT/N_c}
+\frac{1}{6}|\delta_{GS}/2|^{(q_i+q_j+q_k)/2}\lambda_{ijk}Q^iQ^jQ^k,
\nonumber \\
f^{\rm eff}_a&=&kT+\Delta f, \qquad {\cal P}_{\rm eff}={\cal
P}=\mbox{constant}, \eea where $t=T+T^*$ and we made the final
field redefinition $Q^i\rightarrow |\delta_{GS}/2|^{q_i/2}Q^i$. One
can now compute the vacuum values of $T$, $F^T$ and the resulting
soft terms of visible fields using the above effective SUGRA, and
finds the same results as those in (\ref{vacuumvalue}),
(\ref{aucom}) and (\ref{kkltsoft}) for $\partial_T\ln{\cal P}=0$.

\section{Conclusion}

In this paper, we examined the effects of anomalous $U(1)_A$ gauge
symmetry on SUSY breaking while incorporating the stabilization of
the modulus-axion multiplet responsible for the GS anomaly
cancellation mechanism. Since our major concern is the KKLT
stabilization of the GS modulus, we also discussed some features
such as the $D$-type spurion dominance and the sequestering of the
SUSY breaking by red-shifted anti-brane  which is a key element of
the KKLT moduli stabilization. It is noted also that the $U(1)_A$
$D$-term potential can not be an uplifting potential for dS vacuum
in SUSY breaking scenarios with a gravitino mass hierarchically
smaller than the Planck scale.
 In case of the KKLT
stabilization of the GS modulus,
soft terms of visible fields are determined by the GS modulus
mediation, the anomaly mediation and the $U(1)_A$ mediation which
are generically comparable to each other, thereby yielding the
mirage mediation pattern of the superparticle masses at low energy
scale.

\vspace{5mm}
\noindent{\bf Acknowledgments} \vspace{5mm}

We thank A. Casas, T. Kobayashi, K.-I. Okumura
for useful discussions, and particularly Ian Woo Kim for explaining us
some features of the supersymmetry breaking by red-shifted
anti-brane which are presented in this paper.
This work is supported by the KRF Grant funded by the Korean
Government (KRF-2005-201-C00006), the KOSEF Grant (KOSEF
R01-2005-000-10404-0), and the Center for High Energy Physics of
Kyungpook National University.
K.S.J acknowledges also the support of the Spanish Government under
the Becas MAEC-AECI, Curso 2005/06.

\vskip 1cm \noindent {\bf Appendix A. SUSY breaking by red-shifted
anti-brane}

\vskip 0.5cm

 In this appendix, we discuss the red-shift of
the couplings of 4D graviton and gravitino on the world volume of
anti-brane within the framework
of the supersymmetric Randall-Sundrum model on
$S^1/Z_2$ \cite{susyrs}. The bulk action of the model is given by
\bea \label{5daction1} S_{\rm 5D} &=& -\f{1}{2}\int d^4x dy
\sqrt{-G}\, M_5^3 \left\{\,
{R}_5+\bar{\Psi}^i_M\gamma^{MNP}D_N\Psi_{i\,P}
\right.
\nonumber \\
&&-\left.\f{3}{2}k\epsilon
(y)\bar{\Psi}^i_M\gamma^{MN}(\sigma_3)_{ij} \Psi^j_N
-12k^2+\f{\left(\,
\delta(y)-\delta(y-\pi)\,\right)}{\sqrt{G_{55}}}12k\,\right\},
\eea where $R_5$ is the 5D Ricci scalar for the metric $G_{MN}$,
$\Psi^i_M$ ($i=1,2$) are the symplectic Majorana gravitinos, $M_5$
is the 5-dimensional Planck scale, and $k$ is the AdS curvature.
Here we have ignored the graviphoton as it is not relevant for our
 discussion.
The relations between the gravitino kink mass and the brane
cosmological constants are determined by supersymmetry.
 Imposing the standard orbifold boundary
conditions on the 5-bein and 5D gravitino,
one finds that a slice of AdS$_5$ is a solution of the
equations of motion: \beq \label{5dmetric}
ds^2 =e^{-2kL|y|}\eta_{\mu\nu}dx^\mu dx^\nu+L^2dy^2\quad (-\pi\leq
y\leq \pi), \eeq
where $L$ is the orbifold radius.
The corresponding gravitino  zero mode equation is given by
\bea
\partial_y \Psi^{i}_{(0)\mu} + \frac{L}{2}k\epsilon(y) (\sigma_3)^i_{~j}
\gamma_5\Psi^{j}_{(0)\mu} = 0, \nonumber \eea
yielding  the following 4D
graviton and gravitino zero modes: \bea \label{zeromodes}
G_{(0)\mu\nu}(x,y)&=&e^{-2kL|y|}g_{\mu\nu}(x),
\nonumber \\
\Psi^{i=1}_{(0)_\mu}(x,y)&=&e^{-\frac{1}{2}kL|y|}\psi_{\mu L}(x).
\eea

The above form of wavefunctions reflects the
quasi-localization of the 4D graviton and gravitino zero modes
at the UV fixed point $y=0$,  leading to a
red-shift of the zero mode couplings
at the IR fixed point $y=\pi$.
To make an analogy with the KKLT set-up,
let us introduce a brane of 4D AdS SUGRA at $y=0$ and
an anti-brane of non-linearly realized 4D SUGRA at $y=\pi$.
Written in terms of the 4D zero modes  $g_{\mu\nu}(x)$ and $\psi_\mu$, the UV brane action
 is given by
 \bea S_{\rm UV}
\label{UV} &=&\int d^4x \sqrt{-g}\left[\, 3m_{\rm UV}^2M_0^2
-\frac{1}{2}M_0^2R(g) \right. \nonumber \\
&&\left.-\frac{1}{2}\Big(\epsilon^{\mu\nu\rho\sigma}\bar{\psi}_\mu
\gamma_5\gamma_\nu D_\rho{\psi}_\sigma +m_{\rm UV}\bar{\psi}_{\mu
L}\sigma^{\mu\nu}\psi_{\nu R} +{\rm h.c.}\Big)\,\right]. \eea
As for the anti-brane action with non-linearly realized 4D SUGRA at $y=\pi$, let us
choose the unitary gauge of $\xi^\alpha=0$,
where $\xi^\alpha$ is the Goldstino fermion living on the world-volume of anti-brane.
Then using
\bea
\label{redshift}
G_{(0)\mu\nu}(x,\pi)&=&e^{-2\pi kL}g_{\mu\nu}(x),
\nonumber \\
\Psi^{i=1}_{(0)_\mu}(x,\pi)&=&e^{-\pi kL/2}\psi_{\mu L}(x),
\eea
one easily finds that a generic anti-brane action of
$g_{\mu\nu}$ and $\psi_\mu$
can be written as \cite{clark}
\bea \label{IR} S_{\rm IR}&=& \int d^4x
\sqrt{-{g}}\left[\,
-e^{4A}\Lambda_1^4-\frac{1}{2}e^{2A}\Lambda_2^2R({g})
+e^{2A}Z_1\epsilon^{\mu\nu\rho\sigma}\bar{\psi}_\mu\gamma_5{\gamma}_\nu
D_\rho{\psi}_\sigma \right.
\nonumber \\
&&+\,e^{3A}\Big(\Lambda_3\bar{\psi}_{\mu L}{\sigma}^{\mu\nu}
\psi_{\nu R}+\Lambda_4\bar{\psi}_{\mu L}\psi^\mu_R +{\rm h.c.}\Big)
\nonumber \\
&&+\left.e^{2A}\bar{\psi}_\mu\gamma_5{\gamma}_\nu
D_\rho{\psi}_\sigma \Big(Z_2{g}^{\mu\nu}{g}^{\rho\sigma}
+Z_3{g}^{\mu\rho}{g}^{\nu\sigma}+Z_4
{g}^{\mu\sigma}{g}^{\nu\rho}\Big)\right], \eea
where $e^A\equiv e^{-\pi kL}$
and all the coefficients, i.e.
$\Lambda_i$ and $Z_i$ ($i=1,..,4$), are of order
unity in the unit with $M_5=1$.

In fact, adding the brane actions (\ref{UV}) and
(\ref{IR}) to the bulk action  (\ref{5daction1}) makes the solution
(\ref{zeromodes}) unstable.
This problem can be avoided by
introducing a proper mechanism to stabilze the orbifold radius
$L$. In the KKLT compactifications of Type IIB string theory,
such stabilization is achieved by the effects of fluxes.
Generalization of (\ref{5daction1}), (\ref{UV}) and
(\ref{IR}) incorporating the stabilization of the radion $L$ will modify the
wavefunctions  of the graviton and gravitino zero modes, however
still (\ref{zeromodes})  provides a
qualitatively good approximation for the modified wavefunctions as long as
the quasi-localization of zero modes is maintained.
To compensate the {\it negative} vacuum energy density
of the UV brane, the anti-brane should provide a positive
vacuum energy density: $e^{4A}\Lambda_1^4\simeq
3m_{\rm UV}^2M_0^2$, which requires
$e^{2A}\,\sim\,
m_{\rm UV}/M_{0}$ for $M_0\sim \Lambda_1$.
For this value of the warp factor,
the 4D Planck scale and gravitino mass
are given by
\bea
M_{Pl}^2\simeq \frac{M_5^3}{k}+M_0^2,
\quad
m_{3/2}\simeq m_{\rm UV},
\eea
where we have assumed $M_5\sim k\sim M_0$ and
ignored the contributions suppressed by an additional power of $e^{A}$.
Then one finds that SUSY breaking effects due to the terms of
$S_{\rm IR}$  {\it other than} $e^{4A}\Lambda_1^4$ are suppressed by more
powers of $e^A\sim \sqrt{m_{3/2}/M_{Pl}}$
compared to the effects due to the terms in $S_{5D}$ and $S_{\rm UV}$
even when $\Lambda_i$ and $Z_i$ are all of order unity in the
unit with $M_{Pl}=1$. For instance, the gravitino mass from
$S_{\rm IR}$ is of ${\cal O}(e^{3A}M_{Pl})$, while the gravitino mass
from $S_{\rm UV}$ is $m_{3/2}={\cal O}(e^{2A}M_{Pl})$.

\vskip 0.5cm


\end{document}